%% file: main.tex
\documentclass[fleqn,usenatbib]{mnras}

\usepackage{newtxtext,newtxmath}

\usepackage[T1]{fontenc}

\DeclareRobustCommand{\VAN}[3]{#2}
\let\VANthebibliography\thebibliography
\def\thebibliography{\DeclareRobustCommand{\VAN}[3]{##3}\VANthebibliography}

\usepackage{graphicx}	
\usepackage{xspace}	    
\usepackage[dvipsnames]{xcolor}
\usepackage{url}

\title[Bubble sizes around galaxies]{The reionising bubble size distribution around galaxies}

\author[T.-Y. Lu et al.]{Ting-Yi Lu,$^{1,2}$\thanks{E-mail: tingyi-lu@nbi.ku.dk}
Charlotte A. Mason,$^{1,2}$
Anne Hutter,$^{1,2}$
Andrei Mesinger,$^{3}$
Yuxiang Qin,$^{4,5}$
\newauthor
Daniel P. Stark,$^{6}$
and Ryan Endsley $^{7}$
\\
$^{1}$Cosmic Dawn Center (DAWN)\\
$^{2}$Niels Bohr Institute, University of Copenhagen, Jagtvej 128, 2200 Copenhagen N, Denmark\\
$^{3}$Scuola Normale Superiore, Piazza dei Cavalieri 7, I-56126 Pisa, Italy\\
$^{4}$School of Physics, University of Melbourne, Parkville, VIC 3010, Australia\\
$^{5}$ARC Centre of Excellence for All Sky Astrophysics in 3 Dimensions (ASTRO 3D)\\
$^{6}$Steward Observatory, University of Arizona, 933 N Cherry Ave, Tucson, AZ 85721, USA\\
$^{7}$Department of Astronomy, University of Texas, Austin, TX 78712, USA
}

\date{Accepted XXX. Received YYY; in original form ZZZ}

\pubyear{2023}

\begin{document}

\include{macros}

\label{firstpage}
\pagerange{\pageref{firstpage}--\pageref{lastpage}}
\maketitle

\begin{abstract}
Constraining when and how reionisation began is pivotal for understanding when the first galaxies formed. Lyman-alpha (Ly$\alpha$) emission from galaxies is currently our most promising probe of these early stages. At $z>7$ the majority of galaxies detected with Ly$\alpha$ are in candidate overdensities. Here we quantify the probability of these galaxies residing in large ionised bubbles. We create (1.6 Gpc)$^3$ reionising intergalactic medium (IGM) simulations, providing sufficient volume to robustly measure bubble size distributions around UV-bright galaxies and rare overdensities. We find $\MUV \simlt -16$ galaxies and overdensities are $\simgt10-1000\times$ more likely to trace ionised bubbles compared to randomly selected positions. The brightest galaxies and strongest overdensities have bubble size distributions with highest characteristic size and least scatter. We compare two models: gradual reionisation driven by numerous UV-faint galaxies versus more rapid reionisation by rarer brighter galaxies, producing larger bubbles at fixed neutral fraction. We demonstrate that recently observed $z\sim7$ overdensities are highly likely to trace large ionised bubbles, corroborated by their high \lya detection rates. However, the $z\approx8.7$ association of Ly$\alpha$ emitters in EGS and GN-z11, with Ly$\alpha$ at $z=10.6$, are unlikely to trace large bubbles in our fiducial model -- 11\% and 7\% probability of $>1$\,proper Mpc bubbles, respectively. Ly$\alpha$ detections at such high redshifts could be explained by: a less neutral IGM than previously expected; larger ionised regions at fixed neutral fraction; or if intrinsic Ly$\alpha$ flux is unusually strong in these galaxies. We discuss how to test these scenarios with JWST and the prospects for using upcoming wide-area surveys to distinguish between reionisation models.
\end{abstract}




\section{Introduction} \label{sec:intro}

The reionisation of intergalactic hydrogen in the Universe’s first billion years was likely caused by photons emitted from the first galaxies, and is thus intimately linked to their nature \citep[e.g.,][]{Stark2016,Dayal2018,Mesinger2019_book}. Constraining the reionisation process thus enables us to infer properties of these first luminous sources, importantly giving us information about the earliest generations of galaxies which are too faint to observe directly, even with JWST.
In the past decade, substantial progress has been made in measuring the timing of the late stages of reionisation. The electron scattering optical depth to the CMB indicates reionisation was on-going at $z\sim7-8$ \citep{Planck2018} and the attenuation of Lyman-alpha (\lya, 1216\,\AA) photons by neutral hydrogen in the intergalactic medium (IGM) in the spectra of $z\simgt5$ quasars and galaxies implies the IGM was almost entirely ionised by $z\sim5.5-6$ \citep{McGreer2015,Lu2022,Qin2021b,Bosman2022} but that the IGM was significantly neutral (volume-averaged neutral fraction $\xHI \simgt 0.7$) just $\sim300$ Myr earlier at $z\sim8$ \citep[e.g.,][]{Davies2018b,Hoag2019a,Mason2019b, Bolan2022}.

While we are beginning to reach a consensus on when the end stages of reionisation occurred, we still do not understand \textit{how} it happened. Which sources drove it and when did it start? The onset of reionisation provides pivotal information about the onset of star formation. Simulations predict the first reionised regions grow around overdensities \citep[e.g.,][]{Furlanetto2004b,zahn_simulations_2007,Trac2007,Mesinger2007,Ocvirk2018,Hutter2020,qin_dark-ages_2022}, but, while there are strong hints \citep{Castellano2016b,Tilvi2020, Hu2021,endsley_strong_2022,Jung2022, Larson2022}, this is yet to be robustly confirmed observationally.
Furthermore, the ionising emission properties of high-redshift sources are still highly uncertain, and, with current constraints on the reionisation timeline alone, there is a degeneracy between reionisation driven by numerous low mass galaxies with low ionising emissivity (e.g. ionising photon escape fraction $\sim5\%$), and rarer bright galaxies with high ionising emissivity \citep[e.g.,][]{Greig2017c,Mason2019c, Finkelstein2019, Naidu2020}. However, the clustering strength of the dominant source population has a large impact on the expected size distribution of ionised bubbles \citep[e.g.,][]{McQuinn2007b,Mesinger2016,Hassan2018,Seiler2019a}. Thus, identifying and measuring large ionised regions at early times provides vital information about the reionisation process.

Before we will detect the 21-cm power spectrum \citep[e.g.,][]{Pober2014, Morales2010} the most promising tool to study the early stages of reionisation and the morphology of ionised regions is \lya emission from galaxies, which is strongly attenuated by neutral hydrogen \citep[e.g.,][]{Malhotra2006,Stark2010,Dijkstra2014,Mesinger2015,Mason2018}. If reionisation starts in overdensities we expect a strong increase in the clustering of \lya-emitting galaxies (LAEs) in the early stages of reionisation \citep{McQuinn2007a,Sobacchi2015, Hutter2015a}. Strong evidence of enhanced clustering has not yet been detected in wide-area \lya narrow-band surveys \citep[e.g.,][]{Ouchi2017}, likely because these surveys have mostly observed at $z<7$, when the IGM is probably still $<50\%$ neutral \citep[e.g.,][]{Mason2019c, Qin2021b} and thus the clustering signal is expected to be weak \citep{Sobacchi2015}. 

However, spectroscopic studies of $z>7$ galaxies selected by broad- and narrow-band imaging in smaller fields have yielded tantalizing hints of spatial inhomogeneity in the early stages of reionisation. In particular, an unusual sample of four UV luminous ($\MUV \sim -22$) galaxies detected in CANDELS \citep{Koekemoer2011,Grogin2011} fields (three of which are in the EGS field) with strong Spitzer/IRAC excesses, implying strong rest-frame optical emission, were confirmed with \lya emission at $z\approx7.1\,,7.3,\,7.7$, and 8.7 \citep{Zitrin2015a,Oesch2015,Roberts-Borsani2016,Stark2017}. Furthermore, \lya was recently detected at $z=10.6$ in the UV-luminous galaxy GNz11 \citep{Bunker2023}. The high detection rate of \lya in these UV bright galaxies is at odds with expectations from lower redshifts, where UV-faint galaxies are typically more likely to show strong \lya emission \citep[e.g.,][]{Stark2011,Cassata2015}. 

This may imply that these galaxies trace overdensities which reionise early, or that they have enhanced \lya emission due to young stellar populations and hard ionising spectra, or, more likely, a combination of these two effects \citep{Stark2017,Mason2018b, Endsley2021,Roberts-Borsani2022a,Tang2023}. Photometric follow-up around some of these galaxies has found evidence they reside in regions that are $\simgt3\times$ overdense \citep{leonova_prevalence_2022,Tacchella2023}. Furthermore, spectroscopic follow-up for \lya in neighbors of these bright sources has proved remarkably successful: to date, of the $\sim30$ galaxies detected with \lya emission at $z>7$, 14 of these lie within a few physical Mpc of three UV luminous galaxies detected in the CANDELS/EGS field at $z\approx7.3,\,7.7$ and 8.7 \citep{Tilvi2020,Jung2022,Larson2022,Tang2023}. 
Do these galaxies reside in large ionised regions? Due to the high recombination rate at $z\simgt10$ large ionised regions require sustained star formation over $\simgt100$\,Myr \citep[e.g.,][]{Shapiro1987}, thus detection of large ionised regions at early times would imply significant early star formation.

Assessing the likelihood of detecting \lya emitting galaxies during reionisation requires knowledge of the expected distribution of ionised bubble sizes around the observed galaxies. Previous work has focused on predicting the size distribution of all ionised regions during reionisation, as is required for forecasting the 21-cm power spectrum \citep[e.g.,][]{Furlanetto2005,Mesinger2007,Geil2016,Lin2016}. However, as galaxies are expected to be biased tracers of the density field \citep[e.g., ][]{Adelberger1998,Overzier2006,Barone-Nugent2014a}, these ionised bubble size distributions likely underestimate the expected ionised bubble sizes around observable galaxies. The 21-cm galaxy cross-power spectrum for different halo masses \citep[e.g.][]{Lidz2009a, Park2014} reflects the typical ionised region size around different halo masses. However, the size distributions of ionised regions were not discussed in previous works. \citet{Mesinger2008a} show the \lya damping wing optical depth distributions around galaxies of various masses at $z\sim9$, finding the most massive halos have the lowest optical depth with smallest dispersion in optical depth, which corresponds to being hosted by larger bubble sizes with smaller variance in bubbles compared to lower mass halos, though that work did not model the UV magnitude of the halos and only presented optical depths for halos $M_h < 2\times10^{11}\,\Msun$. The correlation between galaxy properties and their host ionised bubbles has been explored in some semi-analytic simulations \citep{Mesinger2008a,Geil2017,Yajima2018,qin_dark-ages_2022}, finding that more luminous galaxies are likely to reside in large ionised bubbles. However, these studies have been restricted to small volumes, (100 cMpc)$^3$, simulations with only a handful of UV-bright galaxies and overdensities, so Poisson noise is large, or models of cosmological Str\"omgren spheres which do not account for the overlap of bubbles \citep{Yajima2018}.

In this paper we create robust predictions for the size distribution of ionised bubbles around observable ($\MUV\lesssim-16$) galaxies. We create large volume (1.6 cGpc)$^3$ simulations of the reionising IGM using the semi-numerical code \cmfast \citep{Mesinger2011}. With these simulations, we can robustly measure the expected bubble size distribution around rare overdensities and UV-bright galaxies ($\MUV\lesssim-22$ or $M_{\mathrm{halo}}\gtrsim10^{11}\Msun$) to compare with observations. We assess how likely the observed $z>7$ associations of \lya emitters are to be in large ionised bubbles, finding that while $z\sim7$ observations are consistent with our current consensus on the reionisation timeline, \lya detections at $z>8$ are very unexpected. We further demonstrate how different reionising source models produce very different predictions for the bubble size distribution at any neutral fraction. We discuss the prospect of using upcoming wide-area surveys to distinguish the reionising source models based on our bubble size distribution predictions by observing a large number of overdensities to chart the growth of the first ionised regions. 

This paper is structured as follows: we describe our simulations in Section~\ref{sec:methods}, we present our results on the bubble size distributions as a function of galaxy luminosity and overdensity, and compare with observations in Section~\ref{sec:results}. We discuss our results in Section~\ref{sec:disc} and conclude in Section~\ref{sec:conc}. We assume a flat $\Lambda$CDM cosmology with $\Omega_m=0.31,\,\Omega_\Lambda=0.69,\,h=0.68$ and magnitudes are in the AB system.

\section{Methods} \label{sec:methods}

In the following sections, we describe our simulation setup and analysis framework. In Section~\ref{sec:methods_sims} we describe our reionisation simulations. In Section~\ref{sec:methods_galaxy} we describe how we populate simulated halos with galaxy properties and in Section~\ref{sec:methods_Rbub} we describe how we measure the ionised bubble size distribution using the mean free path method and the watershed algorithm.

\subsection{Reionisation simulations} \label{sec:methods_sims}

\begin{figure*}
    \centering
    \includegraphics[width=\textwidth]{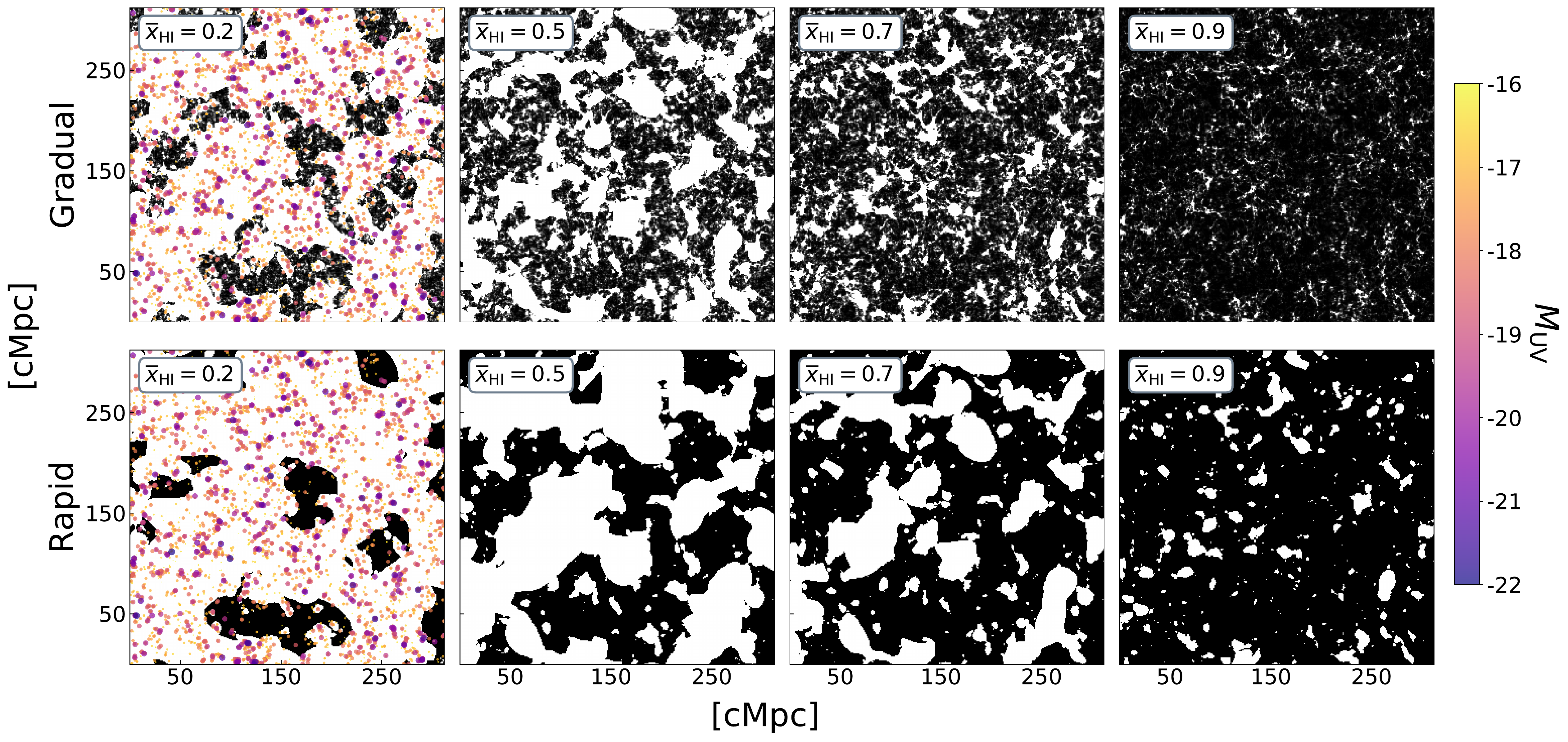}
    \caption{Slices from our simulations at $\xHI=0.2,0.5,0.7,0.9$ for \FG\ (upper panel) and \BG\ (lower panel). White regions show ionised gas and black regions show neutral gas. We show 1.5\,cMpc slices in a $300\times300$\,cMpc region of our (1.6\,cGpc)$^3$ coeval cubes. We plot galaxies in this slice, color-coded by \MUV, in the leftmost column. Here we only show galaxies with $-22\leq\MUV\leq-16$ for demonstration purposes. }
    \label{fig:sim}
\end{figure*}

To study the link between galaxy environment and reionisation, we use the semi-numerical cosmological simulation code, \cmfast v2\footnote{\url{https://github.com/andreimesinger/21cmFAST}} \citep{Mesinger2011,Sobacchi2014a,mesinger_evolution_2016}. \cmfast first creates a 3D linear density field at high redshift, which is evolved to the redshift of interest using linear theory and the Zel'dovich approximation. The ionisation field (and other reionisation observables such as 21cm brightness temperature) is then generated using an excursion-set theory approach assuming an ionisation-density relation and a given reionisation model. In this way, \cmfast can quickly simulate reionisation on large scales ($>100$ Mpc), with a simple, flexible model for the properties of reionising galaxies.

Here we briefly describe the creation of ionisation boxes before proceeding to our simulation setups, and refer the reader to \citet{Mesinger2011,mesinger_evolution_2016} for more details. For a density box at redshift, $z$, a cell (at position \textbf{\textit{x}}) is flagged as ionised if
\begin{equation} \label{eqn:21cmfast_ionised}
    \zeta f_{\rm{coll}}(\textbf{\textit{x}}, z, R, \overline{M}_{\rm min}) \geq 1+\overline{n}_{\rm rec}(\textbf{\textit{x}}, z, R), 
\end{equation}
where $f_{\rm{coll}}(\textbf{\textit{x}}, z, R, \overline{M}_{\rm min})$ is the fraction of a collapsed matter residing in halos larger than a minimum halo mass, $\overline{M}_{\rm min}$, inside a sphere of radius $R$, and $\overline{n}_{\rm rec}$ is the average cumulative number of recombinations. $\zeta$ is an ionising efficiency parameter:
\begin{equation} \label{eqn:zeta_ion}
\zeta=20\left(\frac{N_{\rm \gamma}}{4000}\right)\left(\frac{f_{\rm esc}}{0.1}\right)\left(\frac{f_{\rm \ast}}{0.05}\right)\left(\frac{f_{\rm b}}{1}\right),         
\end{equation}
where $N_{\rm \gamma}$ and $f_{\rm esc}$ are the input reionisation parameters, the number of ionising photons per stellar baryon, and the ionising photon escape fraction, respectively. $f_{\rm \ast}$ is the fraction of galactic gas in stars. $f_{\rm b}$ is the fraction of baryons inside the galaxy. While we expect a variation of these parameters with halo mass and/or time \citep[see e.g.,][]{Wise2014, Kimm2014, Xu2016, Trebitsch2017, Lewis2020, Ma2020}, simply changing $\zeta$ and $\overline{M}_{\rm min}$ can encompass a broad range of scenarios for reionising source models and thus produce different reionising bubble morphologies \citep[e.g.,][]{Mesinger2016}. High values of $\zeta$ and $\overline{M}_{\rm min}$ correspond to reionisation dominated by rare, massive galaxies, which require a larger output of ionising photons to produce a reionisation timeline consistent with observations, while low $\zeta$ and $\overline{M}_{\rm min}$ values correspond to reionisation driven by numerous faint galaxies with weaker ionising emissivity. 

In this paper, we simulate large-scale boxes of dark matter halos and the IGM ionisation field in order to produce robust bubble size distributions as a function of galaxy properties with minimal Poisson noise, using two different reionising source models. We produce (1600\,cMpc)$^{3}$ coeval boxes at $z=[7,8,9,10]$, with a grid size of 1024 pixels, resulting in a resolution of $\sim1.6$\,cMpc. We generate a catalogue of dark matter halos from the density fields associated with these boxes using extended Press-Schechter theory \citep{Sheth2001} and a halo-filtering method (see \citet{Mesinger2007} for full description of the method) which allows us to generate halos with accurate halo mass function down to $\rm M_{\odot}\gtrsim 10^{8}$. We use identical initial conditions (and thus density field and halo catalogue at each redshift) for all of our models, so in our analysis below we can isolate the impact of the reionisation source model on the bubble size distribution in different galaxy environments. 

We create ionisation boxes spanning $\xHI=0.1-0.9$ ($\Delta$\xHI=0.1), using Equation~\ref{eqn:21cmfast_ionised}, for two reionising source models, similar to the approach of \citet{Mesinger2016}, which span the plausible range expected by early galaxies: 
\begin{enumerate}
    \item \textbf{\FG}: Reionisation driven by faint, low mass galaxies down to the atomic cooling limit ($M_\mathrm{min} = 5\times10^8 M_{\odot}$, $\MUV\lesssim-11.0$). Reionisation driven by numerous faint galaxies leads to a gradual reionisation process, where the IGM can begin to reionise very early. We show in Figure~\ref{fig:sim} that the ionised regions in this model start slowly and gradually grow and overlap. We use this as our fiducial model.
    \item \textbf{\BG}: Reionisation driven by rarer bright galaxies ($M_\mathrm{min} = 10^{10} M_{\odot}$, $\MUV\lesssim-19.5$). As massive galaxies take more time to assemble, reionisation starts later and the morphology is characterised by rarer, larger ionised regions at fixed neutral fractions.
\end{enumerate}
For each model, at each redshift, we vary $\zeta$ so as to compare different reionisation morphologies at the same \xHI. In the end, we create a total of 72 simulations: 4 (redshift) $\times$2 (reionisation model) $\times$ 9 (\xHI) ionisation boxes, and 4 (redshift) halo catalogues. In addition, in Sec.~\ref{sec:res_interp}, to compare our simulations with observations, we expand the \xHI range at the high-\xHI end to \xHI=[0.85,0.90,0.95] at $z=9$ for the two models.

Example slices of the ionisation field from the two sets of simulations are shown in Figure~\ref{fig:sim}. This clearly shows that the \BG\ model has larger, rarer bubbles compared to the \FG\ model at fixed \xHI. Underdense regions are more likely to be ionised in the \FG\ model. This is because in the \FG\ model, faint galaxies, which live in a wider density range, are able to ionise the IGM. While in the \BG\ model, only bright, more massive galaxies, which most likely only live in overdensities, can ionise the IGM.

\begin{figure}
    \centering
    \includegraphics[width=\columnwidth]{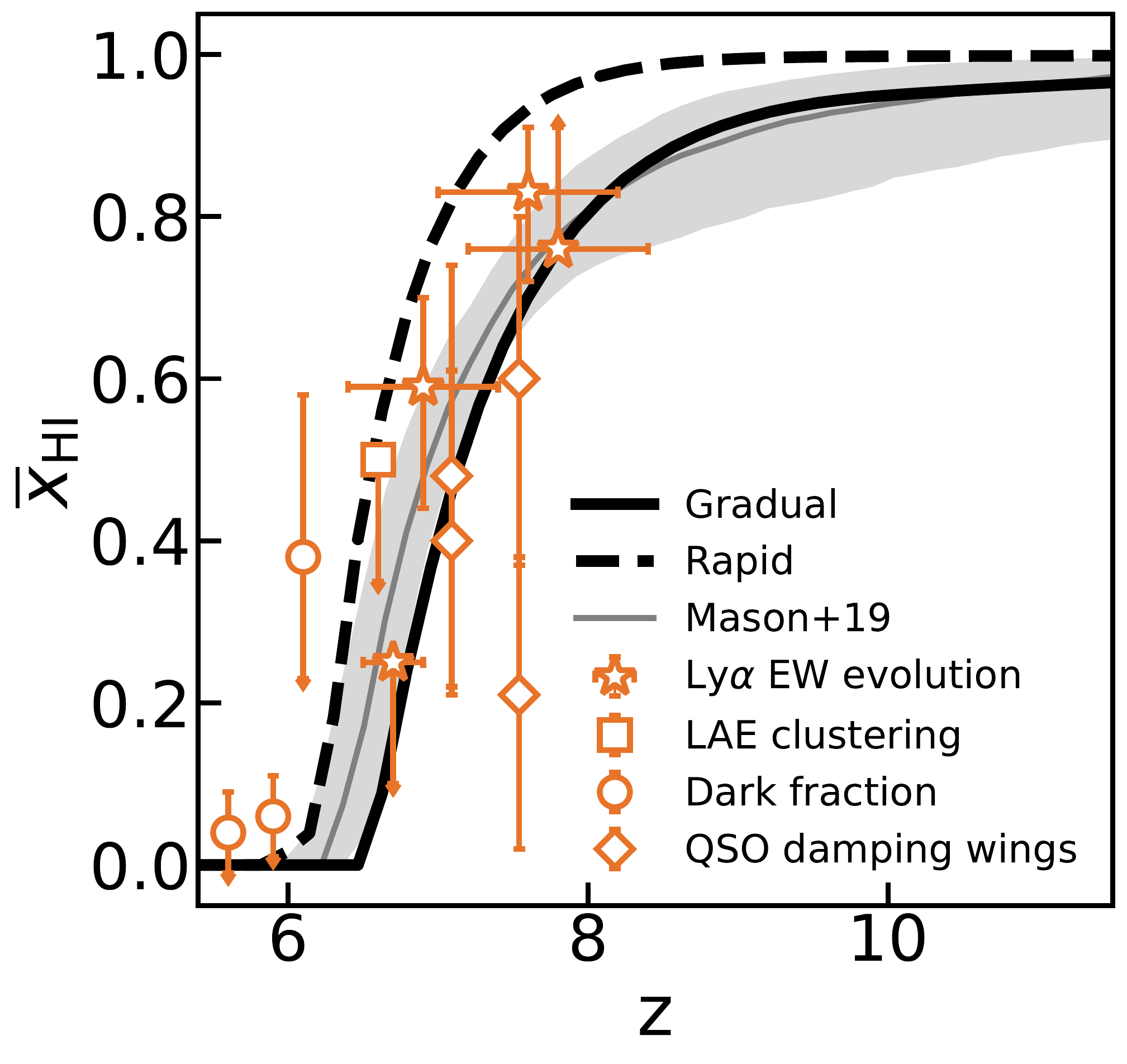}
    \caption{Example reionisation timelines for the \FG\ model (solid line) and the \BG\ model (dashed line) for demonstration purposes. Different symbols are neutral fractions constrained by \lya equivalent width \citep[stars,][]{Mason2018, Mason2019c,Bolan2022}, \lya emitter clustering \citep[squares][]{Sobacchi2015}, \lya forest dark pixels fraction \citep[circles,][]{McGreer2015}, and QSO damping wings \citep[diamonds,][]{Davies2018, Greig2019} observations. The grey line with shaded region is the reionisation timeline and its 16-84 percentile inferred using the aforementioned observations \citep{Mason2019c}. In the following we will use this grey posterior for \xHI for comparing to observations as a function of redshift, the \BG\ and \FG\ models are shown just to illustrate how these models differ when the ionisation efficiency is fixed (see Section~\ref{sec:methods_sims}). }
    \label{fig:xHI_demo}
\end{figure}

Figure~\ref{fig:xHI_demo} shows potential reionisation timelines of the two reionisation models, for demonstration purposes only. To produce example reionisation histories for our two models we follow the standard procedure \citep[e.g.,][]{Robertson2010} and generate an ionising emissivity from the product of the halo mass density, integrated down to the two mass limits described above, and an ionising efficiency, $\zeta$. We alter $\zeta$ for both models to fix the redshift of the end of the reionisation to $z\sim6$. The \FG\ model has an earlier onset of reionisation and slower redshift evolution of \xHI compared to the \BG\ model. We note that as we use coeval boxes we do not assume a model reionisation history in this work, rather we will use non-parametric reionisation timeline inferred by \citet{Mason2019b} from independent constraints on the IGM neutral fraction, including the \lya equivalent width distribution \citep{Mason2018, Mason2019c, Hoag2019a}, \lya emitter clustering \citep[][]{Sobacchi2015}, \lya forest dark pixels fraction \citep[][]{McGreer2015}, and QSO damping wings \citep[][]{Davies2018, Greig2019} and the \citet{Planck2018} electron scattering optical depth. 

\subsection{Galaxy population model} \label{sec:methods_galaxy}

To populate halos with realistic galaxy properties we use a conditional UV luminosity to halo mass relation, to assign UV luminosities, with intrinsic scatter, to our halo catalogue. We follow \citet{Ren2019} and assume UV magnitudes at a given halo mass are drawn from a Gaussian distribution with dispersion $\sigma$ and median $M_{\textsc{uv}, c}(M_\text{h}, \sigma, z)$:
\begin{equation} \label{eqn:Muv_Mh_scatter}
p(M_\textsc{uv} \mid M_\text{h}) = \frac{1}{\sqrt{2\pi}\sigma} \exp \left(\frac{-[M_\textsc{uv} - M_{\textsc{uv}, c}(M_\text{h}, \sigma, z)]^2}{2\sigma^2}\right).
\end{equation}
The dispersion was originally introduced to explain scatter in the Tully--Fisher relation \citep{Yang2005}. It is a free parameter in our model, and following \citet{Whitler2020} we assume $\sigma=0.5$\,mag. \citet{Ren2019} found that this value is consistent with observed luminosity functions over $z\sim6-10$, and this value is also consistent with the expected variance due to halo assembly times \citep{Ren2018,Mason2022}. \citet{Whitler2020} found that this scatter has only a minor impact on the transmission of \lya from galaxies in the reionising IGM, so we do not expect it to significantly change the relationship between galaxy luminosity and the size of the ionised bubbles they reside in.

The median relation $M_{\textsc{uv}, c}(M_\text{h}, \sigma, z)$ is set by calibration to the UV luminosity function. \citet{Ren2019} showed that above $M_h \simgt 10^{12}\Msun$ a flattening is required in $M_{\textsc{uv}, c}(M_\text{h})$ to maintain consistency with the observed UV LFs -- which can be thought of as a critical mass or luminosity threshold for star formation. Given that our halo catalogue contains only a small number (0.001$\%$ of the total catalogue) of $>10^{12}\Msun$ halos at $z\sim7$, and far fewer at $z>7$ due to the steepness of the halo mass function, we do not consider this flattening. We thus use the $M_{\textsc{uv}, c}(M_\text{h}, z)$ relations from the \citet{Mason2015} UV luminosity function model as the median UV magnitudes for Equation~\ref{eqn:Muv_Mh_scatter}. Our resulting luminosity functions are consistent with $z\sim7-10$ observations over the range where observations are currently magnitude complete: $-22 \simlt \MUV \simlt -17$ \citep[e.g.,][see Appendix~\ref{app:LF}]{Bouwens2021}.

\subsection{Measuring bubble sizes} \label{sec:methods_Rbub}

We measure the size of ionised regions, \Rbub, using both the mean-free-path (MFP) method \citep{Mesinger2007} and the watershed algorithm \citep{vincent1991watersheds}, an image segmentation algorithm which was first applied to reionisation simulations by \citet{Lin2016}. 

\citet{Lin2016} tested a range of approaches for estimating the sizes of ionised bubbles in simulations and determined these two methods were optimal compared to other techniques in the literature because they most accurately recover input ionised bubble size distributions, can account for overlapping bubbles, and produce sizes corresponding to a physically intuitive quantity.
Other commonly used approaches for modelling the bubble size distribution, i.e. the excursion set formulation \citep{Furlanetto2004b,Furlanetto2005} or approaches which grow cosmological Stromgren spheres around halos \citep[e.g.,][]{Yajima2018} will underestimate the largest bubble sizes because these approaches do not include the effect of overlapping bubbles.

Here we describe these two methods, and their advantages and limitations. We will discuss how our resulting bubble size distributions compare to works using other methods in Section~\ref{sec:disc_compare}.

\subsubsection{Mean free path (MFP)} \label{sec:methods_Rbub.MFP}

This method was first used to measure ionised bubble sizes by \citet{Mesinger2007}. It is essentially a Monte-Carlo ray-tracing algorithm, which enables us to measure a probability distribution for ionised bubble sizes by estimating the distance photons travel before they encounter neutral gas. We randomly choose a starting position (or the position of a galaxy, as described later), if the cell is fully ionised, we measure the distance from that position to where we encounter the first neutral or partially ionised cell at a random direction. Given our simulation resolution, the smallest bubble size we can measure is $\sim1$\,cMpc. If the position is neutral, we set $\Rbub=0$\,cMpc. We measure bubble sizes over the full simulation volume by sampling the distance to neutral gas from $10^5$ random positions and sightlines to build bubble size distributions for our simulations.

In Section~\ref{sec:results_BSD_MUV} we will show the bubble size distribution as a function of galaxy \MUV, to estimate the sizes of ionised bubbles around observable galaxies. For this, we use the mean free path method as defined above, but start our measurements at the position of each galaxy in the simulation box. 

We also will measure the bubble size distribution as a function of galaxy overdensity to compare with current observations. Galaxy overdensity depends on the dark matter density of the underlying field \citep[e.g.][]{Cole1989, Mo1997, Sheth2001}: $n = \overline{n} (1 + b\delta)$,
where $n$ is the number density of galaxies observed in a field, $\overline{n}$ is the mean cosmic number density, $b$ is the bias, and $\delta$ is the dark matter density in the field. Since \cmfast populates halos and calculates $x_{\mathrm{HI}}$ based on galaxy number density via the excursion set formulation \citep{Furlanetto2004b}, we expect a strong relation between bubble size and galaxy overdensity \citep[e.g.,][]{Mesinger2007}.

We define the observed overdensity, \od, as the number, $N$, of galaxies brighter than a given limit in a survey volume relative to the number expected in that volume based on the average in the whole simulation box, $\langle N \rangle$. To measure overdensity using our galaxy catalogue, described in Section~\ref{sec:methods_galaxy}, for a mock survey, we discard galaxies with $\MUV > M_{\rm UV,lim}$, where $M_{\rm UV,lim}$ is the UV magnitude limit in an observed overdensity. While the galaxy catalogue and \xHI boxes are generated from the same density field, as described in Section~\ref{sec:methods_sims}, galaxies are given sub-grid positions, thus to compare the overdensity and \xHI fields we convert the resulting galaxy catalogue into a galaxy number count grid of the same size as the \xHI grids. Then we convolve the galaxy number count grid with a 3-D kernel of the survey volume and divide the value of each cell by $\langle N \rangle$, the mean number count per cMpc$^{3}$ in the halo box, to obtain the overdensity in each cell. 

Cells in the resulting overdensity box correspond to positions with a given overdensity above the magnitude limit within the volume. We can then carry out an analogous procedure to that described above using the mean free path algorithm to find the bubble size distribution as a function of overdensity using the mean free path method, by starting in positions of a given overdensity.

\subsubsection{Watershed algorithm} \label{sec:methods_Rbub.watershed}

This method was first used to measure ionised bubble sizes in reionisation simulations by \citet{Lin2016}. It is an image segmentation algorithm which treats constant values of a scalar field as contour lines corresponding to depth in a tomographic map, which it then ``floods'' to break up the images into separate water basins \citep{vincent1991watersheds}.

We use the implementation of the watershed algorithm in \texttt{skikit-image} \citep{scikit-image}. We apply the algorithm to 3D binary \xHI cubes. We first apply the `distance transform' to calculate the Euclidean distance, $d_i$ of every point to the nearest neutral region (if the point is neutral then the distance is zero). We invert the distances to `depths': $d_i \rightarrow -d_i$. Centres of bubbles are then local minima in the depth cube and the bubble boundaries are identified by flooding regions starting from the local minima, and marking where regions meet -- these are contours of constant depth $d_i$.

As with any image segmentation algorithm, the identification of local minima will lead to over-segmentation, as every local minimum will be marked as a unique bubble, even if it is overlapping with a larger one, thus a threshold must be used to avoid this. We follow the prescription of \citet{Lin2016} and use the `H-minima transform' to essentially `fill in' small basins. We identify basins with a relative depth of $h$ from the local minimum to the bubble boundary and set $d_i \rightarrow d_i + h$ for these regions, reducing the depth of the local minima. After the H-minima transform, we can again identify bubbles as above and see that large bubbles are correctly identified. This process may remove small isolated bubbles which had depth $< h$. These can be added back in manually using the initial segmentation cube.

The H-minima threshold $h$ is a free parameter, we use $h=2.5$, which is fixed so that the resulting bubble size distribution is comparable to that obtained with the MFP method above and bubbles do not suffer too much from over-segmentation. We solve for the value of $h$ by minimising the Kullback-Leibler divergence \citep[KL divergence;][]{Kullback1968} between the watershed bubble size distribution and MFP bubble distribution in a (500\,cMpc)$^3$ sub-volume of our simulation. We obtain a cube with the cells corresponding to unique bubbles labelled. From this we can calculate the volume of each bubble and calculate the size as the radius of each bubble assuming they are spherical: $R = (3V/4\pi)^{1/3}$

The watershed algorithm is a more computationally intensive method than the MFP method, and requires some tuning of the $h$ threshold, so we predominantly use the MFP approach. However, the watershed algorithm has a significant advantage in that it can measure the absolute number of bubbles in a volume. It is also possible to use it to directly connect galaxies and their host bubbles. We will use it in Section~\ref{sec:res_forecasts} to make forecasts for the number of large bubbles expected in upcoming wide-area surveys.

\section{Results} \label{sec:results}

Previous works have focused on simulating the global bubble size distribution, in order to produce predictions for 21-cm experiments \citep[e.g.][]{Furlanetto2005, Mesinger2007,Geil2016,Lin2016}. Some 21cm-galaxy cross correlation studies \citep[e.g.][]{Lidz2009a,Park2014} calculate the correlation scales for various halo masses but do not directly calculate the bubble size distribution. Here we focus on the expected bubble size distribution \textit{around observable galaxies}, which are likely to be more biased density tracers, and thus we expect are likely to trace the largest bubbles.

In Section~\ref{sec:results_BSD_MUV} we present the bubble size distribution as a function of galaxy UV luminosity, and in Section~\ref{sec:results_BSD_overdensity} we show the bubble size distribution as a function of galaxy overdensity. The impact of different reionising source models on the bubble size distribution is discussed in Section~\ref{sec:results_BSD_model}. We demonstrate in Appendix~\ref{app:z} that our results do not significantly depend on redshift. In Section~\ref{sec:res_interp} we use our simulations to interpret recent observations of \lya emission in overdensities at $z\simgt7$, and we make predictions for upcoming wide-area observations in Section~\ref{sec:res_forecasts}.

\subsection{Bubble size distribution as a function of UV luminosity} 
\label{sec:results_BSD_MUV}

\begin{figure*}
    \centering
    \includegraphics[width=0.8\textwidth]{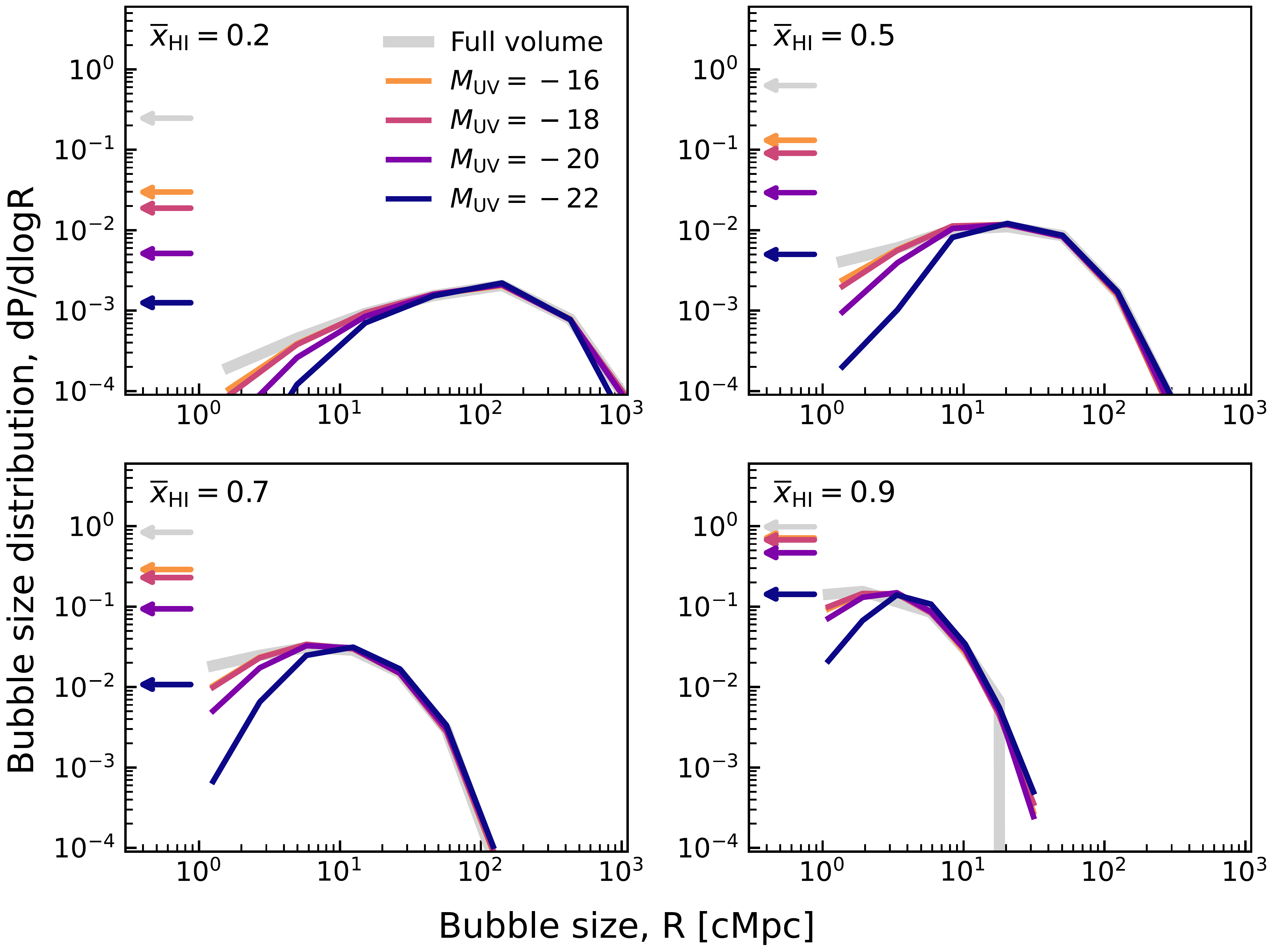}
    \caption{Bubble size distributions as a function of UV luminosity for $\MUV = -16, -18, -20, -22$. We also show the bubble size distribution from the full volume as a thick grey line in each simulation. The fractions of galaxies in $R<0.8$\,cMpc bubbles (below our resolution limit) or neutral cells are marked with arrows. Each panel shows a different volume-averaged IGM neutral fraction, \xHI. As the neutral fraction decreases, the bubble size distributions shift to higher values, as expected as bubbles grow as reionisation progresses. With increasing UV luminosity, the probability that a galaxy resides in big bubbles increases.}
    \label{fig:pR_Muv}
\end{figure*}

To first order, UV luminosity traces dark matter halo mass and thus density \citep[e.g.,][]{Cooray2005a, Tempel2009, Mason2015}. We thus expect the brightest galaxies to reside in the most massive halos in overdense regions, and therefore these galaxies are likely to sit in large bubbles which reionised early.

We quantify this in Figure~\ref{fig:pR_Muv}, where we show the size distribution of ionised bubbles around galaxies of a given UV luminosity as a function of the volume-averaged IGM neutral fraction, \xHI, in our simulations, compared to the bubble size distribution in the full volume. This is essentially the distribution at the mean density, $\delta=0$. We measure the distribution of bubble sizes in 4 \MUV bins: $\MUV=-16, -18, -20, -22$, with $\Delta\MUV=0.1$. We show our fiducial \FG\ simulation but will compare it to the \BG\ simulation in Section~\ref{sec:results_BSD_model}. 

In contrast to previous literature we also include the fraction of galaxies (or randomly selected pixels for our full volume bubble size distribution) which are in neutral regions in our simulation. We mark these fractions with arrows in Figure~\ref{fig:pR_Muv}. These sources may reside in ionised bubbles below our resolution limit ($\sim1$\,cMpc for bubble radius).  Including these occurrences in our bubble size distribution leads to important insights about the environments of galaxies as we discuss below. We note that the `full volume' bubble size distribution excluding neutral cells and those below our resolution limit is equivalent to the bubble size distributions presented in previous literature \citep[e.g.,][]{Furlanetto2005, Mesinger2007}.

Figure~\ref{fig:pR_Muv} shows that as \xHI decreases, the bubble size distributions shift to higher values, which is expected as ionised regions grow. Compared to the bubble size distribution in the full volume, we see three important features of the bubble size distributions which we describe below.

First, while the bubble size distribution in the full volume has a high fraction of bubbles with $R\lesssim1$\,cMpc, observable galaxies ($\MUV\lesssim-16$) are $>10-1000\times$ more likely to be in bubbles rather than neutral regions. This is because galaxies are biased tracers of the density field and therefore trace ionised regions more closely. At the end stages of reionisation, $\xHI \simlt0.5$, we find only $\simlt10\%$ of observable galaxies are in small ionised or neutral regions below our resolution limit. This is consistent with the idea of the `post-overlap' phase of reionisation \citep[e.g.,][]{Miralda-Escude2000}, where the majority of galaxies lie within ionised regions and only voids remain to be ionised.

We see a strong trend with UV luminosity, where the brightest galaxies are always least likely to be in small ionised or neutral regions, while UV-faint galaxies have a more bimodal bubble size distribution. The proportion of UV-faint galaxies in small ionised or neutral regions is high early in reionisation: but declines rapidly from $\sim60\%$ at $\xHI\sim0.9$ to $\sim10\%$ at $\xHI\sim0.5$ for $\MUV\simgt-18$ galaxies. This is driven by the clustering properties of the UV-faint galaxies as we discuss below.
This may explain the low detection rate of \lya in UV-faint galaxies at $z\simgt8$ \citep{Hoag2019a,Mason2019c,Morishita2022b} compared to the higher detection rate in UV bright galaxies seen by \cite{Jung2022} at the same redshift. 

Second, we see that the bubble size distribution around observable galaxies peaks at a similar size for all \MUV bins, which indicates that, on average, these galaxies are in the same bubbles. This peak, corresponding to
 the mean size of ionised regions, has been described as a `characteristic' scale, $R_\mathrm{char}$ \citep[e.g.,][]{Furlanetto2005}. In the following we refer to the mean size of ionised regions as the characteristic size. We see that the characteristic scale of ionised regions increases by over two orders of magnitude during reionisation\footnote{Note that our characteristic scale is at least an order of magnitude higher than that presented by \citet{Furlanetto2005} due to our use of the mean free path approximation, which captures the sizes of overlapping bubbles \citep{Lin2016}}. However, we do find an increasing characteristic scale as a function of UV luminosity: galaxies brighter than $\MUV \simlt -20$ are expected to reside in bubbles $\sim1.5-2\times$ larger than the characteristic bubble scale in the full volume.

Finally, we see that the width of the bubble size distribution decreases as galaxy UV luminosity increases.
This is due to the clustering of galaxies: UV bright galaxies are more likely to be in overdense regions which will reionise early, whereas UV faint galaxies can be both `satellites' in overdense, large ionised regions, or `field galaxies' in  less dense regions which remain neutral for longer \citep[see also][]{Hutter2017, Hutter2020, qin_dark-ages_2022}
This figure demonstrates that UV-faint galaxies will have very significant sightline variance in their \lya optical depth, and highlights the importance of using realistic bubble size distributions for inference of the IGM neutral fraction \citep[see also, e.g.][]{Mesinger2008,Mason2018b}.

\subsection{Bubble size distribution as a function of galaxy overdensity}
\label{sec:results_BSD_overdensity}

\begin{figure*}
    \centering
    \includegraphics[width=0.8\textwidth]{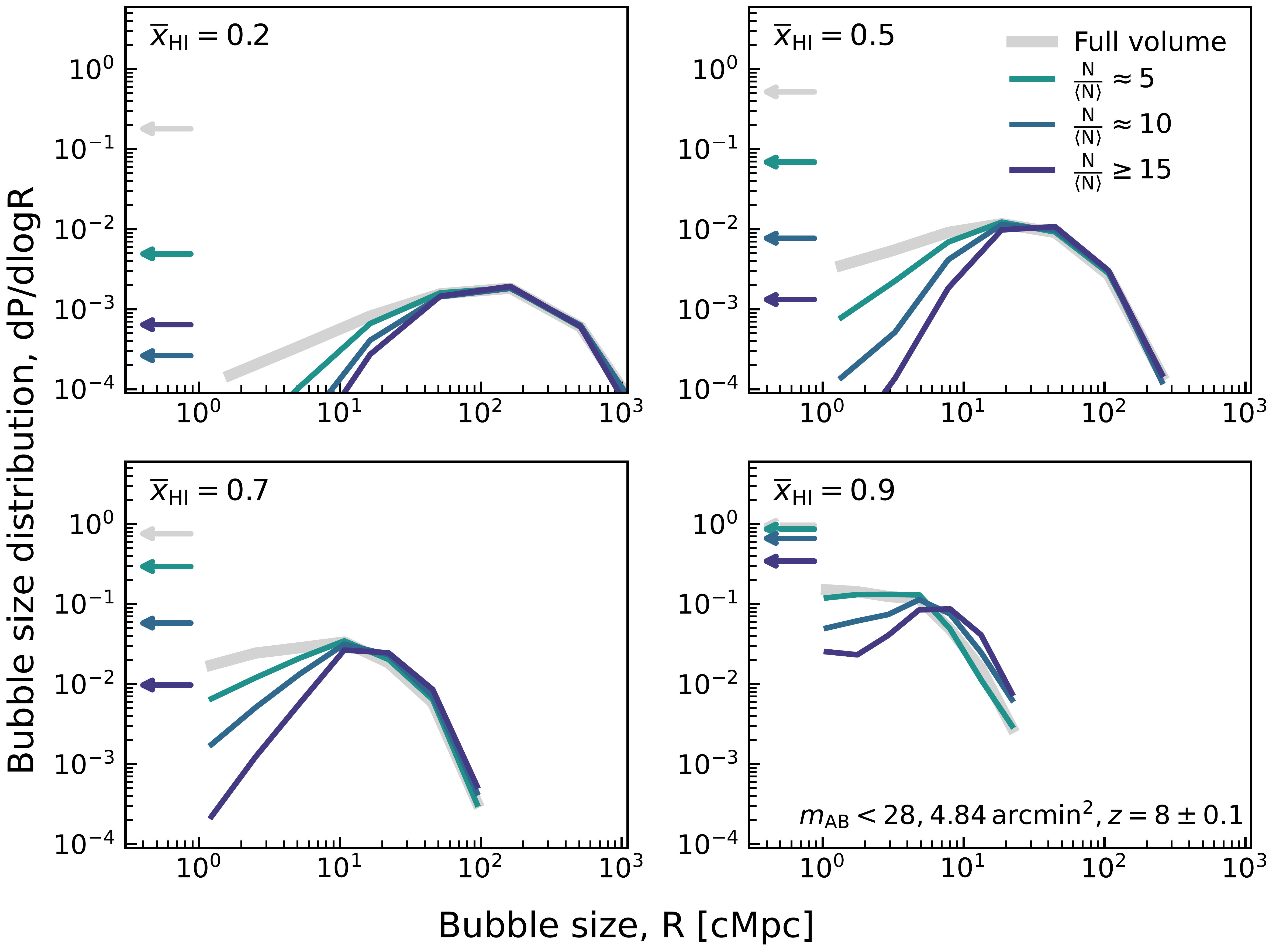}
    \caption{Bubble size distributions as a function of galaxy overdensity at $z=8\pm0.1$ in a 4.84 arcmin$^{2}$ area ($\sim \left(13 \mathrm{cMpc}\right)^{3}$) with a survey limit of $m_{\mathrm AB}=28$, for $\od \approx 5,10$ and $\geq15$, where $\langle N\rangle=0.84$. We also show the bubble size distribution from the full volume as a thick grey line in each simulation. The fractions of galaxies in $R<0.8$\,cMpc bubbles (below our resolution limit) or neutral cells are marked with arrows. Each panel shows a different \xHI. More overdense regions host larger \Rbub early at high \xHI. As \xHI decreases, \Rbub of less overdense regions begins to catch up and ends up having similar bubble size distribution to those of the most overdense regions. This agrees with the general reionisation picture, that overdense regions are reionised first.}
    \label{fig:PR_differentodxIH}
\end{figure*}

In this section, we investigate the distribution of bubble sizes as a function of galaxy overdensity, \od. This distribution should directly
reflect how structure formation affects reionisation.

As described in Section~\ref{sec:methods_Rbub} an observed galaxy overdensity, \od, depends on the survey depth and volume. For our investigation here we explore expected overdensities within a medium-deep JWST observation within 1 NIRISS pointing (or $1/2$ of the NIRCam field-of-view), aiming to simulate observations similar to those obtained by the JWST/NIRISS pure-parallel PASSAGE survey \citep{Malkan2021}. We thus use a survey limiting depth of $m_\mathrm{AB}=28$ and area 4.84 sq. arcmin with a redshift window of $\Delta z = 0.2$. This corresponds to [$\MUV_\mathrm{,lim}$, $V_\mathrm{survey}$]=[-19, 2014\,cMpc$^{3}$] at $z=8\pm0.1$. We follow the procedure described in Section~\ref{sec:methods_Rbub} to create a cube of \od using these survey parameters, and then select 200,000 cells\footnote{Due to the sampling variance and slightly different binning, the bubble size distributions for the full volume here and in Section~\ref{sec:results_BSD_MUV} are slightly different.} to measure the bubble size distribution as a function of overdensity.

Figure~\ref{fig:PR_differentodxIH} shows the bubble size distributions for  $\od\approx5$, $\od\approx10$, and  $\od\gtrsim15$, along with the bubble size distribution in the full volume as a function of \xHI, for the \FG\ model. As in Section~\ref{sec:results_BSD_MUV} we see the clear trend that the bubble size distributions increase to higher values as the universe reionises, but we can now identify where the reionisation process begins. We can see that the most overdense regions reionise first and inhabit the largest ionised bubbles. As in Section~\ref{sec:results_BSD_MUV}, we investigate three clear trends in the bubble size distribution as a function of galaxy overdensity.

First, overdense regions start and finish carving out ionised bubbles earlier compared to regions at the mean density. We see a much larger proportion of overdense regions already in $\Rbub>1$\,cMpc bubbles early in reionisation. We find at $\xHI=0.9$, when only 10\% of the total IGM volume is ionised, $\simgt 30\%$ of the $\od\geq10$ regions are already in $\Rbub>1$\,cMpc bubbles. By $\xHI=0.5$, all of the $\od\geq10$ regions are in $\Rbub>1$\,cMpc bubble. This demonstrates that early in reionisation, we expect only the strongest overdensities to trace large ionised regions. 

Second, ionised bubbles around overdense regions are larger than the characteristic bubble size in the full volume, particularly in the early stages of reionisation.
At $\xHI=0.8$, the characteristic bubble size around $\od\geq10$ regions is $\Rbub\sim10$\,cMpc, which is $\sim2\times$ larger than the mean bubble size in the full volume at that time, and large enough for significant \lya transmission \citep[e.g.,][]{Miralda-Escude1998b,Mason2020,qin_dark-ages_2022}. Detection of \lya in a highly neutral universe is thus not unexpected if the LAEs are in highly overdense regions.

The mean bubble size of overdense regions grow more slowly than that of less overdense regions. In the early stage of reionisation, bubbles around the most overdense regions grow in isolation and do not merge with similarly sized bubbles because most overdense regions are far away from each other. By contrast, bubbles created by less overdense regions are more likely to grow rapidly by merging with other bubbles. 

Finally, again we see the bubble size distributions are broad, but that the strong overdensities have the narrowest distribution of bubble sizes because they are guaranteed to trace ionised environments, whereas less dense regions can be isolated, and therefore in smaller bubbles, or contained within large scale overdensities in large bubbles.

\subsection{Bubble size distribution as a function of reionising source model}
\label{sec:results_BSD_model}

\begin{figure*}
    \centering
    \includegraphics[width=0.7\textwidth]{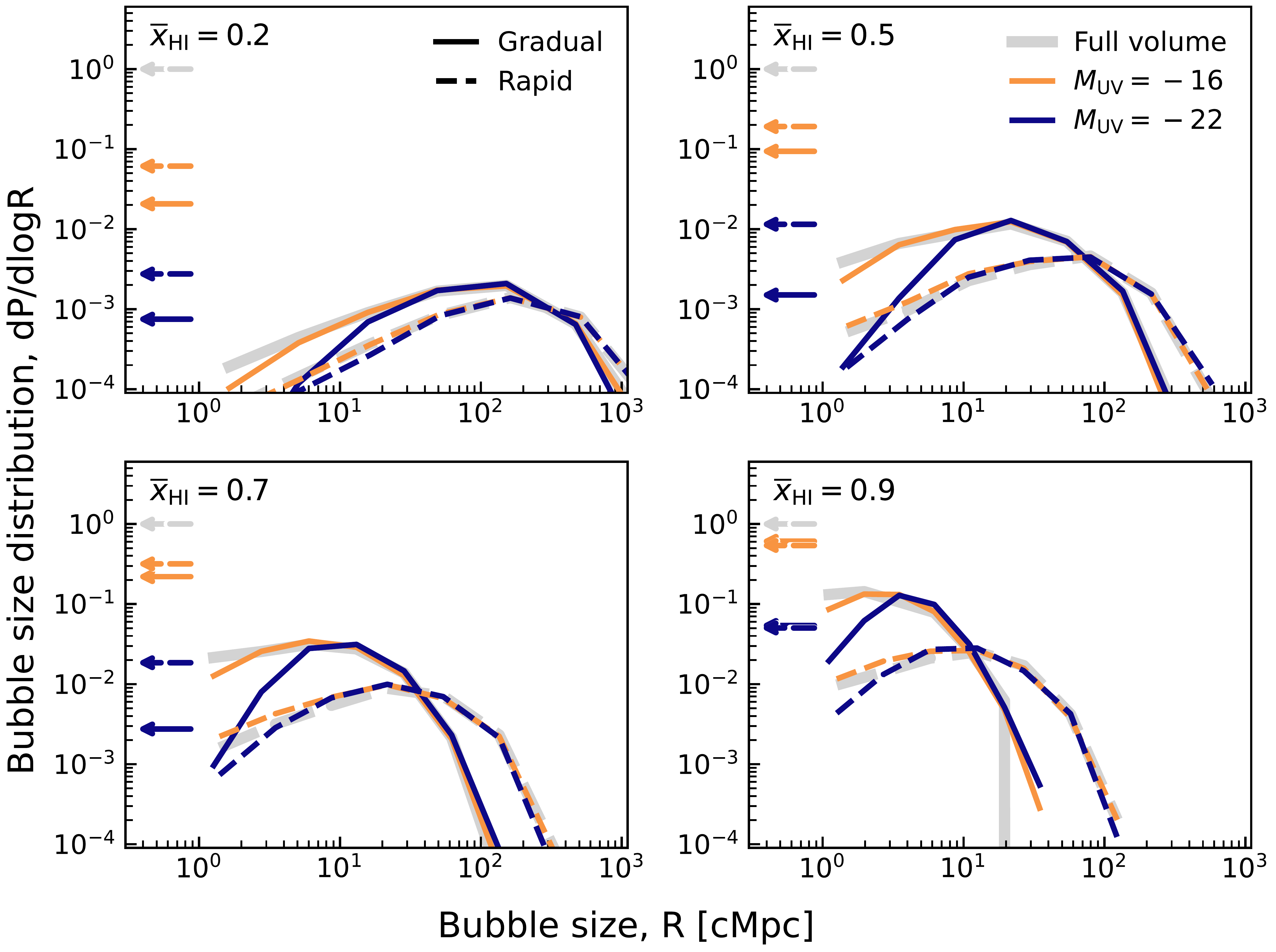}
    \caption{Bubble size distributions as a function of UV luminosity for $\MUV = -16, -22$, for the \FG\ (solid lines) and \BG\ (dashed lines) reionisation models. We also show the bubble size distribution in the full volume as a thick grey line for each simulation. The fractions of galaxies in $R<0.8$\,cMpc bubbles (below our resolution limit) or neutral cells are marked with arrows. Each panel shows a different volume-averaged IGM neutral fraction. We see that the bubble size distributions are broader for the \BG\ models than for the \FG\ model at $\xHI\simgt 0.5$. The bubble size distributions of the \BG\ model peak at $\Rbub\gtrsim10$cMpc since as early as $\xHI=0.9$. By contrast the distributions of \FG\ models start with $\Rbub\lesssim6$\,cMpc at $
    \xHI=0.9$, and gradually evolve to converge with  the \BG\ models as IGM becomes more ionised.}
    \label{fig:pR_Muv_BG_FG}
\end{figure*}
\begin{figure*}
    \centering
    \includegraphics[width=0.7\textwidth]{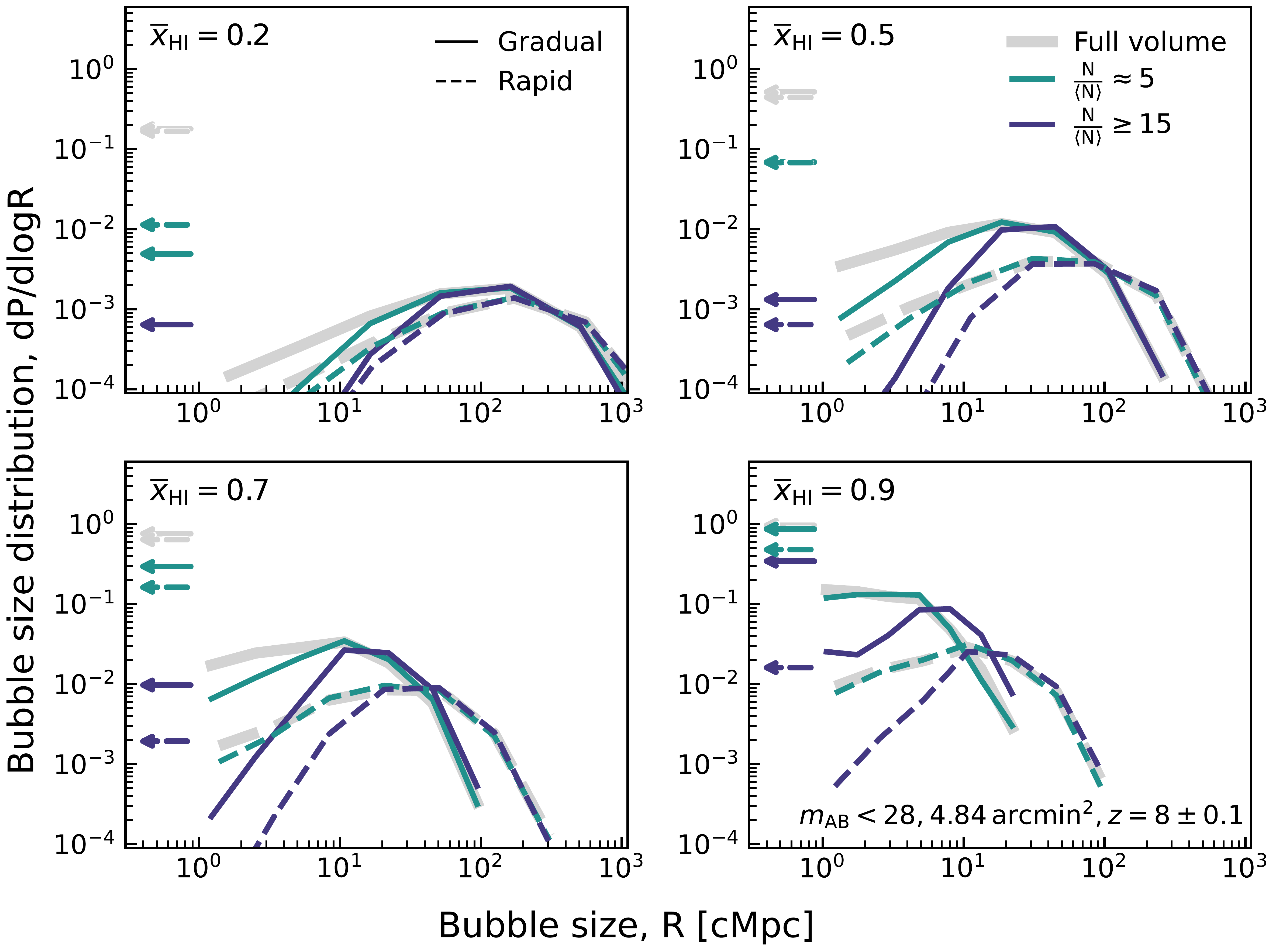}
    \caption{Bubble size distributions as a function of overdensity for $\od = 5, >15$, for the \FG\ (solid lines) and \BG\ (dashed lines) reionising source models. We also show the total bubble size distribution as a thick grey line in each simulation. The fractions of galaxies in $R<0.8$\,cMpc bubbles (below our resolution limit) or neutral cells are marked with arrows. Each panel shows a different volume-averaged IGM neutral fraction. In the \BG\ model we see that bubble size distribution of $\od>7$ already shows little bimodality at $\xHI=0.9$. Galaxies in $\od>15$ regions are mostly in bubbles of $\Rbub>7$. In contrast, in \FG\ model even galaxies in $\od>5$ regions are in bubbles of $\Rbub<7$.}
    \label{fig:pR_od_BG_FG}
\end{figure*}

In the previous sections we have shown the bubble size distribution using only our fiducial \FG\, faint-galaxies driven, reionisation model. Here we demonstrate how the bubble size distribution changes if instead reionisation is driven by rarer, brighter galaxies in our \BG\ model. We show the bubble size distributions for the two models as a function of the IGM neutral fraction in Figures~\ref{fig:pR_Muv_BG_FG} and ~\ref{fig:pR_od_BG_FG} for galaxies of given \MUV and galaxy overdensities. 

Both models have qualitatively similar bubble size distributions but the \BG\ model predicts much large bubble sizes at fixed neutral fraction, particularly at the earliest stages of reionisation. A key prediction of the \BG\ model is the existence of large ($\sim30-100$\,cMpc) bubbles at the earliest stages of reionisation, $\xHI\sim0.9$, in order to fill the same volume with ionised hydrogen around the more biased ionising sources.

First, galaxies in the \FG\ model are more likely to reside in neutral IGM at the beginning of reionisation, compared to galaxies in the \BG\ model.
At $\xHI=0.9$, $\sim80\%$ of the $\MUV=-16$ galaxies have bubble sizes no greater than $1$\,cMpc in the \FG\ model. By contrast, in the \BG\ model, only $\sim60\%$ of $\MUV=-16$ galaxies are in such neutral regions at the same \xHI. At the mid-point of reionisation ($\xHI=0.5$), UV-faint galaxies ($\MUV=-16$) in the \FG\ model ($\sim9\%$) are half as likely to be in small ionised/neutral regions compared to UV-faint galaxies in the \BG\ model ($\sim20\%$). This is because the early ionised regions in the \BG\ model are concentrated around the most overdense regions, compared to a more uniform coverage of bubbles seen in the \FG\ model (see Figure~\ref{fig:sim}). 
In the \BG\ model isolated faint galaxies cannot create $\Rbub>1$\,cMpc bubbles around themselves, because reionisation is dominated by $\MUV \simlt 19.5$ galaxies in this model. Therefore isolated faint galaxies remain in $\Rbub<1$\,cMpc bubbles even at the mid-point of the reionisation.

Second, galaxies in the \BG\ model blow out big ionised bubbles early in the reionisation. However, the bubble sizes do not grow as rapidly as those in the \FG\ model. 
At the beginning of reionisation, the characteristic bubble size
in the \BG\ model is $\Rchar\approx10$\,cMpc. In the \FG\ model, the characteristic bubble size $\sim3\times$ smaller:
no more than 3\,cMpc. 
By the late stages of reionisation ($\xHI=0.1$), the mean bubble sizes in the \BG\ model are $\sim300$\,cMpc. However, in the \FG\ model, the mean bubble size has grown twice as rapidly, reaching $\sim200$\,cMpc. 
The different evolutionary trends reflect the different bubble-merging histories of the two models. In the \FG\ scenario, many faint galaxies create small ionised bubbles and soon merge together to form big bubbles. In the \BG\ model, big bubbles form early, however, but due to the rarity of bright ionising galaxies, bubbles are less likely to merge and immediately double in size compared to those in the \FG\ model. 

We can see from this comparison that there can be a degeneracy between the \FG\ and \BG\ model. If we find evidence of a large ($>10$\,cMpc) bubble at high redshift, it could be explained by a bright-galaxies-driven reionisation at a high neutral fraction, or by the faint-galaxies-driven reionisation but with a lower neutral fraction. However, independent information on the reionisation history and/or information from the dispersion of bubble sizes along multiple sightlines could break this degeneracy. We discuss this in Section~\ref{sec:disc_reionisation}.

\subsection{Interpretation of current observations}
\label{sec:res_interp}

\begin{table*}
\caption{Assumed properties of $z\simgt7$ associations of \lya emitters used in our simulations.}
\label{tab:bubbles}      
\centering      
\setlength{\tabcolsep}{4pt} 
\begin{tabular}{ lcccccccccc }
\hline\hline
\multicolumn{7}{c}{} & \multicolumn{2}{c}{$p(R > 1\,\mathrm{pMpc})^\ast$} & \multicolumn{2}{c}{} \\ \cline{8-9} 
Field & $z$ & $N_\mathrm{LAEs}$ & Minimum \MUV & \xHI$^\dagger$ & Volume$^\ddagger$ [pMpc$^{3}$] & Overdensity & Full volume & In overdensity & $R_\mathrm{char}$\,$^\ast$ [pMpc] & References \\
\hline\hline
COSMOS     & 6.8 & 9 & $-20.4$ & $0.44^{+0.09}_{-0.17}$ & 140  & $>3$   & 0.52 (0.72) & 0.93 (0.99) &6.4 (11.2) &  [1] \\
BDF        & 7.0 & 3 & $-19.5$ &$0.56^{+0.09}_{-0.08}$ & 53 & $>3$  & 0.27 (0.57) & 0.84 (0.98) &3.1 (6.3) &  [2-5] \\
EGS        & 7.7 & 7 & $-19.5$ & $0.76^{+0.05}_{-0.09}$ & 2.6 & $>3$  & 0.07 (0.26) & 0.39 (0.67) &1.0 (2.2) &  [6-10]\\
Abell 2744 & 7.9 & 0 & $-17.7$ & $0.80^{+0.06}_{-0.09}$ & 0.001 &$>130$& 0.10 (0.29) & 0.27 (0.53) &0.5 (1.6) & [11,12] \\
EGS        & 8.7 & 2 & $-19.5$ & $0.93^{+0.02}_{-0.15}$ & 12 & $>3$  & 0.01 (0.07) & 0.11 (0.42) &0.8 (1.6) &  [8,9,13,14]\\
GOODS-N    &10.6 & 1 & $-18.6$ & $>0.92$ & 2.6 &$>24$  & 0.004 (0.02) & 0.07 (0.48) &0.5 (1.1) &  [15-17]\\
\hline
\end{tabular}

$^\dagger$ Using the non-parametric reionisation history posteriors by \citet{Mason2019c} including constraints from the CMB optical depth, quasar dark pixel fraction and measurements of the \lya damping wing in quasars and galaxies. We calculate $p(R > 1\,\mathrm{pMpc})$ by marginalising $p(R > 1\,\mathrm{pMpc} | \xHI)$ over the \xHI posterior at each redshift inferred by \citet{Mason2019c}. $^\ddagger$We assume a redshift window of $\Delta z = 0.2$ except for the COSMOS and Abell 2744 regions which are spectroscopically confirmed. For those regions we use the volumes estimated by \citet{endsley_strong_2022} and \citet{Morishita2022b} respectively.\\
$^\ast$ Calculated using \FG\, (\BG).
[1] \citet{endsley_strong_2022}, [2] \citet{Vanzella2011}, [3] \citet{Castellano2016b}, [4] \citet{Castellano2018}, [5] \citet{Castellano2022}, [6] \citet{Oesch2015}, [7] \citet{Tilvi2020}, [8] \citet{leonova_prevalence_2022}, [9] \citet{Tang2023}, [10] \citet{Jung2022}, [11] \citet{Morishita2022b}, [12] \citet{Ishigaki2016}, [13] \citet{Zitrin2015a}, [14] \citet{Larson2022}, [15] \citet{Oesch2016}, [16] \citet{Bunker2023}, [17] \citet{Tacchella2023}.
%
\end{table*}

\begin{figure}
    \centering
    \includegraphics[width=\columnwidth]{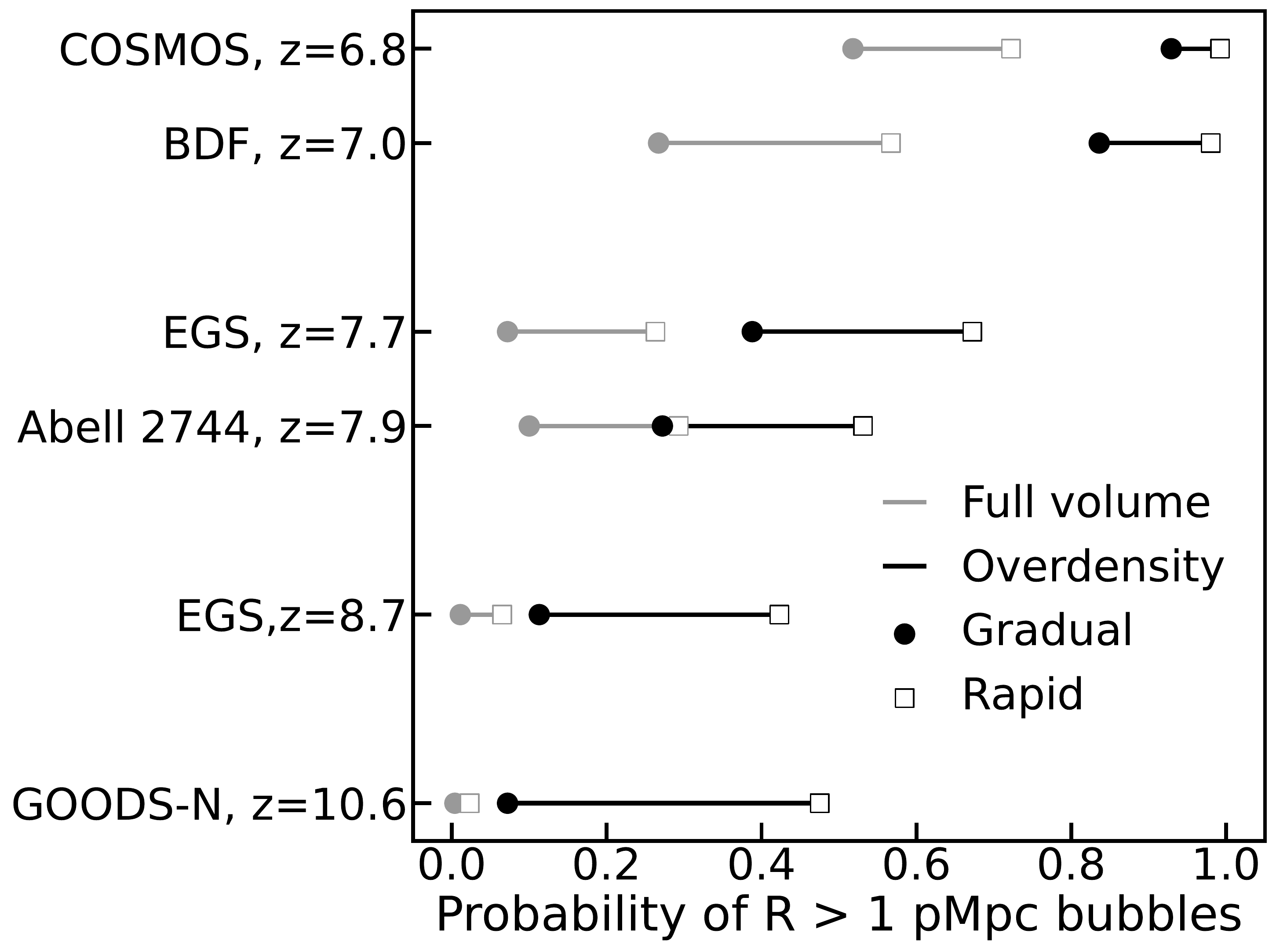}
    \caption{Probability in our models of finding a bubble size $>1$\,pMpc around regions similarly overdense to the observed $z\simgt7$ associations of \lya emitters in our \FG\ and \BG\ simulations (black lines). Grey lines show the range of probabilities in the full simulation volume. We use the IGM neutral fractions expected at these redshifts \citep{Mason2019b}. The plot is discussed in Section~\ref{sec:res_interp} and a summary of our simulation setup is given in Table~\ref{tab:bubbles}. The bubble size distributions for all these fields is shown in Figure~\ref{fig:BSD_obs_all}. It is highly likely for the observed $z\sim7$ \lya emitting galaxies to reside in large ionised bubbles. At $z\gtrsim8$, large bubbles are unexpected even in overdensities. }
    \label{fig:BSD_obs}
\end{figure}

In this section, we use our simulations to interpret some recent observations of \lya emission at $z\simgt7$ in candidate overdensities. Here we aim to establish if the enhanced \lya visibility in these regions can be explained by the sources tracing an ionised overdensity, and how likely that scenario is given our consensus timeline of reionisation and either of our two reionisation models.

We focus on observations of $z\simgt7$ \lya emission from galaxies in 6 regions in the sky, in candidate overdensities: the COSMOS field at $z\approx6.8$ \citep{endsley_strong_2022}, BDF field at $z\approx7.0$ \citep{Vanzella2011,Castellano2016b, Castellano2018, Castellano2022}, EGS field with two regions at $z\approx7.7$ and 8.7 \citep{Oesch2015, Roberts-Borsani2016, Tilvi2020, leonova_prevalence_2022, Larson2022, Jung2022, Tang2023}, the field behind the galaxy cluster Abell 2744 at $z\approx7.9$ \citep{Morishita2022b} and the area around the $z=10.6$ galaxy GNz11 \citep{Oesch2016,Bunker2023,Tacchella2023}.

To compare with observations at known redshifts, we will switch from comoving to proper distance units. Due to the incompleteness of the observations, here we aim to create bubble size distributions for regions in our simulations that are approximately similar to those observed. Our simulations are coeval boxes at $z=7,8,9,10$, so in the following we use the box closest in redshift to the observations, but use the observed redshift to fix assumed IGM neutral fractions and to calculate physical distances. We demonstrate in Appendix~\ref{app:z} that our bubble size distributions do not depend significantly on redshift.

We create mock observations assuming the same area as the observed overdensities. For the regions which only have photometric overdensities, due to the large redshift uncertainties in the photometric overdensities, but motivated by the $\Delta z \simlt 0.2$ redshift separation of \lya emitters in all of these regions, we assume a redshift window of $\Delta z = 0.2$ for our mock observations. This corresponds to $\sim5-8$\,pMpc at $z\sim7-10$. In all cases we will assume the observed overdensity of Lyman-break galaxies to be the same in that smaller volume as in the true observed volume. This means we are likely overestimating the true overdensity, in that case our estimated probabilities can be seen as upper limits.

Using these assumed volumes we then use the method described in Section~\ref{sec:methods_Rbub.MFP} to convolve our galaxy field with the volume and depth kernel of the observations to create a cube of overdensity matched to each observation setup. We then select regions in the overdensity cube which match the observed overdensity estimates. 

We assess the probability of the observed overdensities lying in ionised regions $>1$\,pMpc in radius, which would allow $\simgt30\%$ of \lya flux to be transmitted at the rest frame \lya line centre \citep[up to $\sim50\%$ transmission for emission 500\,km s$^{-1}$ redward of linecentre, e.g.,][]{Mason2020,qin_dark-ages_2022}. While the true \lya detection rate will depend on the flux limit of the survey and the \lya flux emitted by the galaxies, before attenuation in the IGM, this threshold gives us a qualitative approach with which to interpret the observations. We defer full forward-modelling of \lya\ observations to a future work. At the redshift of each observation we assume a non-parametric estimate of the IGM neutral fraction, \xHI, inferred by \citet{Mason2019b} described in Section~\ref{sec:methods_sims}. We then calculate the final bubble size distribution by marginalising the bubble size distribution at each \xHI over the inferred \xHI distribution. We also measure the characteristic bubble sizes, $\Rchar$, for the observed overdensities, which indicates the mean size of ionised regions above our resolution limit around the overdensities.

We present a summary of our simulation setups to compare to these observations in Table~\ref{tab:bubbles}, the resulting probability of each region residing in a large ionised region in Figure~\ref{fig:BSD_obs}, and \Rchar. The full bubble size distributions are described in Appendix~\ref{app:obs}.

\subsubsection{$z\sim7$ overdensities in COSMOS and BDF fields}
\label{sec:res_interp_z7}

In the COSMOS field, \cite{endsley_strong_2022} detected \lya in 9/10 $\MUV \lesssim -20.4$, $z\approx6.8$ galaxies in a 140\,pMpc$^{3}$ volume. Using these spectroscopic confirmations, they estimate the lower limit of the overdensity of this region is $\gtrsim3$. They estimate that individual galaxies in this field can create ionised bubbles $\Rbub\sim0.69-1.13$\,pMpc. Taking into account the $\od\sim 3$ overdensity and the ionising contribution from $\MUV<-17$ galaxies, they estimate an ionised bubble radius of $\Rbub\sim3$\,pMpc in this volume. 

We predict that almost 100\% of regions this overdense at $\xHI\approx0.5$ are in $>1$\,pMpc bubbles and the characteristic bubble size is $\Rchar=6.4$\,pMpc at $\xHI\approx0.4$. The high LAE fraction detected by \cite{endsley_strong_2022} is thus consistent with being a typical ionised region in our \FG\ model. In the \BG\ model, we predict even larger bubble sizes around this overdensity: the characteristic bubble size is $\Rchar=11.2$\,pMpc, thus high \lya transmission would also be expected. In both cases we would expect an excess of \lya detections in neighbouring UV-faint galaxies.

In the BDF field, \citet{Vanzella2011} and \cite{Castellano2018} detected 3 $z\sim7.0$ LAEs, with $\MUV=[-21.1,-20.4,-20.4]$. Two of the galaxies have a projected separation of only 91.3 pkpc, and the third is 1.9 pMpc away \citep{Castellano2018}. The photometric overdensity of $z\sim7$ Lyman-break galaxies within $\sim3.86$\,arcmin$^{2}$ around these galaxies is $3-4\times$ times higher than expected \citep{Castellano2016b}.  
Based on the star formation rate and age of the galaxies, and assuming a uniform IGM $\xHI=0.5$ surrounding the sources, \citet{Castellano2018} estimated individual bubble sizes of the two galaxies at $\sim2$\,pMpc separation to be $\Rbub<0.8$\,pMpc. We use a 56\,arcmin$^{2}$ survey area \cite[corresponding to the BDF field, where the sources have an angular separation of 6 arcmin,][]{Vanzella2011} at $z=7\pm0.1$, which is $>3\times$ overdense \citep{Castellano2016b}. We see in Figure~\ref{fig:BSD_obs} that we expect nearly all regions ($\sim84\%$ in our fiducial \FG\ model) with this galaxy overdensity to be inside $>1$\,pMpc ionised bubbles, enabling significant \lya escape. 

The non-detection of \lya with equivalent width $>25$\,\AA\ in twelve surrounding UV-faint galaxies in this region may thus be surprising, but could be explained by a number of reasons, as discussed by \citet{Castellano2018}. For example, even given the predicted most likely bubble size of 4\,pMpc the fraction of transmitted \lya flux may be only $\sim60\%$ for galaxies at the centre of the bubble \citep{Mason2020}, thus with deeper spectroscopy \lya may be detected. It could also be possible that infalling neutral gas in this region resonantly scatters \lya photons emitted redward of systemic \citep[which look blue in the rest-frame of the infalling gas, e.g.,][]{Santos2004a,Weinberger2018,Park2021}. As UV-faint galaxies are likely to be low mass, and thus have a lower HI column density in the ISM compared to UV-bright galaxies, they may emit more of their \lya close to systemic redshift, making it more easily susceptible to scattering by infalling gas. Measurements of systemic redshifts for the galaxies in this region may help explain the complex \lya visibility. Furthermore, given the large photometric redshift uncertainties of the \citet{Castellano2016b} sample, it could also be possible that the actual galaxy overdensity associated with the three LAEs is smaller, thus the expected bubble size is smaller and the faint galaxies may not lie in the same bubble as the detected LAEs.
\subsubsection{$z\sim8$ overdensities in EGS and Abell 2744 fields}
\label{sec:res_interp_z8}

The EGS field contains the majority of $z>7$ LAEs that have been detected to-date \citep{Zitrin2015,Oesch2015,Roberts-Borsani2016,Tilvi2020,Larson2022,Jung2022,Tang2023}. Among these LAEs, \citet{Tilvi2020}, \cite{Jung2022} and \citet{Tang2023} have reported a total of 8 $\MUV<-20$ $z\approx7.7$ LAEs, including the $\MUV=-22$ LAE detected by \citet{Oesch2015}, within a circle of radius $\approx 1$\,pMpc. \citet{Jung2022} estimates the $\Rbub<1.1$\,pMpc for the individual galaxies based on the model of \citet{Yajima2018} which relates \lya luminosity and bubble size. The photometric overdensity around these LAEs has been estimated to be $\od\sim3-5$ \citep{leonova_prevalence_2022}.

We calculate the bubble size distributions using a setup similar to the results of \citet{leonova_prevalence_2022}: an area of 4.5 arcmin$^{2}$ at $z=7.7\pm0.1$, with a limiting UV magnitude $\MUV>-19.5$. The result is shown in Figure~\ref{fig:BSD_obs}, assuming the neutral fraction $\xHI(z=7.7)=0.76_{-0.09}^{+0.05}$ \citep{Mason2019b}. We find $\sim40\%$ of regions this overdense are in large ionised bubbles in the \FG\ model and 70\% of regions in the \BG\ model. We conclude this region is likely consistent with our consensus picture of reionisation.

In the Abell 2744 field, \citet{Morishita2022b} found no \lya detections of 7 $z\approx7.89$, $\MUV>-20$ galaxies. These galaxies are within a circle of radius $\sim$60\,pkpc. This area is $\od\sim130$ overdense for galaxies with $\MUV > -17.5$ \citep{Ishigaki2016}. \citet{Morishita2022b} estimated bubble sizes of $\Rbub\sim0.07-0.76$ pMpc for individual galaxies, based on their ionising properties derived from rest-frame optical spectroscopy with NIRSpec. We generate the bubble size distributions for a region of $>130\times$ overdensity of $\MUV \simlt -17.5$ galaxies within a volume of (0.9\,cMpc)$^3$. 
A bubble size of $\Rbub\sim1$pMpc or larger is unexpected for regions as overdense as this in our \FG\ model at $\xHI\sim0.8$: we find $p(R>1\mathrm{pMpc})=0.27$.

The redshifts of sources in the EGS and Abell2744 fields are very similar. However, \lya has only been detected in the EGS field. We can see in Figure~\ref{fig:BSD_obs} and Table~\ref{tab:bubbles} that our predicted bubble size distributions for EGS are shifted towards higher bubble sizes than in Abell 2744. Although the Abell 2744 region is overdense in UV-faint galaxies, the volume of this region is very small, thus there may not be sufficient ionising emissivity to produce a large-scale ionised region. Thus non-detection of \lya in this overdensity is not surprising.

\subsubsection{$z\sim9-11$ overdensities in EGS and GOODS-N fields}
\label{sec:res_interp_z9}

The highest redshift association of LAEs in the EGS field is a pair at $z\approx8.7$ \citep{Zitrin2015a,Larson2022}, which lies $\sim4$\,pMpc apart. The photometric overdensity around these LAEs has been estimated to be $\od\sim3-5$ \citep{leonova_prevalence_2022}.
We calculate the bubble size distributions using a setup similar to the results of \citet{leonova_prevalence_2022}: an area of 27 arcmin$^{2}$ (corresponding to $\sim6$ HST/WFC3 pointings between the two sources) with $\Delta z = 0.2$, with a limiting UV magnitude $\MUV>-19.5$. 

At $z=8.7$ the inferred IGM neutral fraction is $\xHI=0.93_{-0.15}^{+0.02}$. We predict the probability of finding LAEs at $\xHI\approx0.9$ should be extremely low: in the full simulation volume in our fiducial \FG\ model, we obtain $p(R>1\mathrm{pMpc})=0.01$ and there is $<0.2$\% probability of finding a bubble with $\Rion>4$\,pMpc. Around regions as overdense as that observed we find $p(R>1\mathrm{pMpc})=0.11$. Thus in our fiducial model, we find it is extremely unlikely that the $z\approx8.7$ LAE pair in the EGS field are in one large ionised region. 

The visibility of \lya\ therefore implies some missing aspect in our understanding of this system. If $\xHI$ is lower, there will be a higher chance to find LAEs: we obtain $p(R>1\mathrm{pMpc})=0.17$ in regions this overdense if $\xHI=0.8$, so \xHI will need to be substantially lower to find a high probability of large ionised regions. Alternatively, in the \BG\ model, $p(R>1\mathrm{pMpc})=0.42$ for such an overdensity, and the bubble size distributions at $\xHI=0.8-0.6$ peak at $\Rbub\gtrsim3$\,pMpc: the two LAEs could be in one large ionised bubble. Finally, the \lya\ visibility of these galaxies could be boosted by high intrinsic \lya production as suggested by their other strong emission lines, and potential contribution of AGN \citep{Stark2017,Tang2023,Larson2023}, and facilitated transmission in the IGM if the \lya flux is emitted redward of systemic \citep[e.g.,][]{Dijkstra2011,Mason2018b}.

Finally, \citet{Bunker2023} have detected \lya in GN-z11 at $z=10.6$, in the GOODS-N field \citep{Oesch2016}. 9 fainter galaxy candidates ($m_{\rm AB}\approx29$) at similar redshift are found within a (10 cMpc)$^{2}$ square centred at GN-z11 \citep{Tacchella2023}. We estimate the overdensity of $m_{\rm AB}<29$ ($\MUV<-18.6$), $z=10\pm 0.1$ galaxies in this field using our $z=10$ UV LF, finding that this region is $\sim23\times$ overdense. We obtain $p(R>1\mathrm{pMpc})=0.07$ and 0.48 in the \FG\ and the \BG\ model, respectively. It is thus extremely unlikely that all of the $z\sim11$ galaxies are in one $R>1\mathrm{pMpc}$ ionised region that allows significant \lya transmission, in our fiducial \FG\ model. 

We find $\Rchar=0.5$ and 1.1 pMpc in the \FG\ and the \BG\ model, respectively. The characteristic bubble size is slightly smaller than the largest distance of galaxies from GN-z11 in this field ($\sim0.6$\,pMpc) estimated by \citet{Tacchella2023} from photometric redshifts, implying that most of these galaxies could reside in the same (small) ionised region. 

In summary, our simulations demonstrate that the regions discussed above at $z\sim7$ are extremely likely ($>90\%$) to be in large ionised bubbles, given their large estimated overdensities. We also find it likely ($\simgt40\%$) that the EGS region at $z\approx7.7$ is in a large ionised bubble. However, at higher redshifts we find it very unlikely that the $z\approx8.7$ \lya-emitters in EGS and the $z\approx10.6$ galaxies in GOODS-N, including GNz11, are in large ionised regions ($\sim11\%$ and $\sim7\%$ respectively) in our fiducial \FG\ model. 

If the actual overdensities of these regions are smaller than the photometrically estimated values, we will find it even more unlikely for these galaxies to reside in large ionised bubbles, strengthening our result.
These results clearly demonstrate the importance of measuring the IGM neutral fraction at $z\simgt8$ and distinguishing between reionising source models \citep[e.g.,][]{Bruton2023}, and of understanding intrinsic \lya production and escape in the ISM in these galaxies \citep{Roberts-Borsani2022a,Tang2023}.

\subsection{Forecasts for future observations}
\label{sec:res_forecasts}

Anticipating upcoming large area surveys at $z\simgt7$ we make forecasts for the expected number of large bubbles in the JWST COSMOS-Web survey \citep{Casey2022}, the Euclid Deep survey \citep{Euclid_wide2022, Euclid_deepz6_2022}, and the Roman High-Latitude Survey \citep{Wang2022}. These surveys will detect tens of thousands of UV-bright $z>7$ galaxy candidates which could be used to constrain the underlying density field and pinpoint early nodes of reionisation.

The identification of ionised regions in these large surveys has the potential to distinguish between reionisation models. 
We sample simulated volumes equivalent to the survey areas (0.6\,sq. degrees for COSMOS-Web, 53\,sq. degrees for Euclid-Deep) at $z = 8\pm0.1$ and use the watershed algorithm (Section~\ref{sec:methods_Rbub.watershed}) to identify individual bubbles in these volumes. As we describe below the bubble size distribution in a Euclid Deep-like survey volume will suffer minimal cosmic variance, thus our Euclid forecast can be rescaled to forecast for the the Roman High-Latitude Survey (2000\,sq. deg).
We show in Appendix~\ref{app:z} that the expected bubble sizes, in comoving units, do not depend strongly on redshift, so our results can be easily shifted to other redshifts without expecting significant differences. As discussed above, to reduce over-segmentation we use the H-minima threshold when calculating the bubble sizes using the watershed algorithm, this sets an effective resolution of 3\,cMpc.

\begin{figure}
    \centering
    \includegraphics[width=\columnwidth]{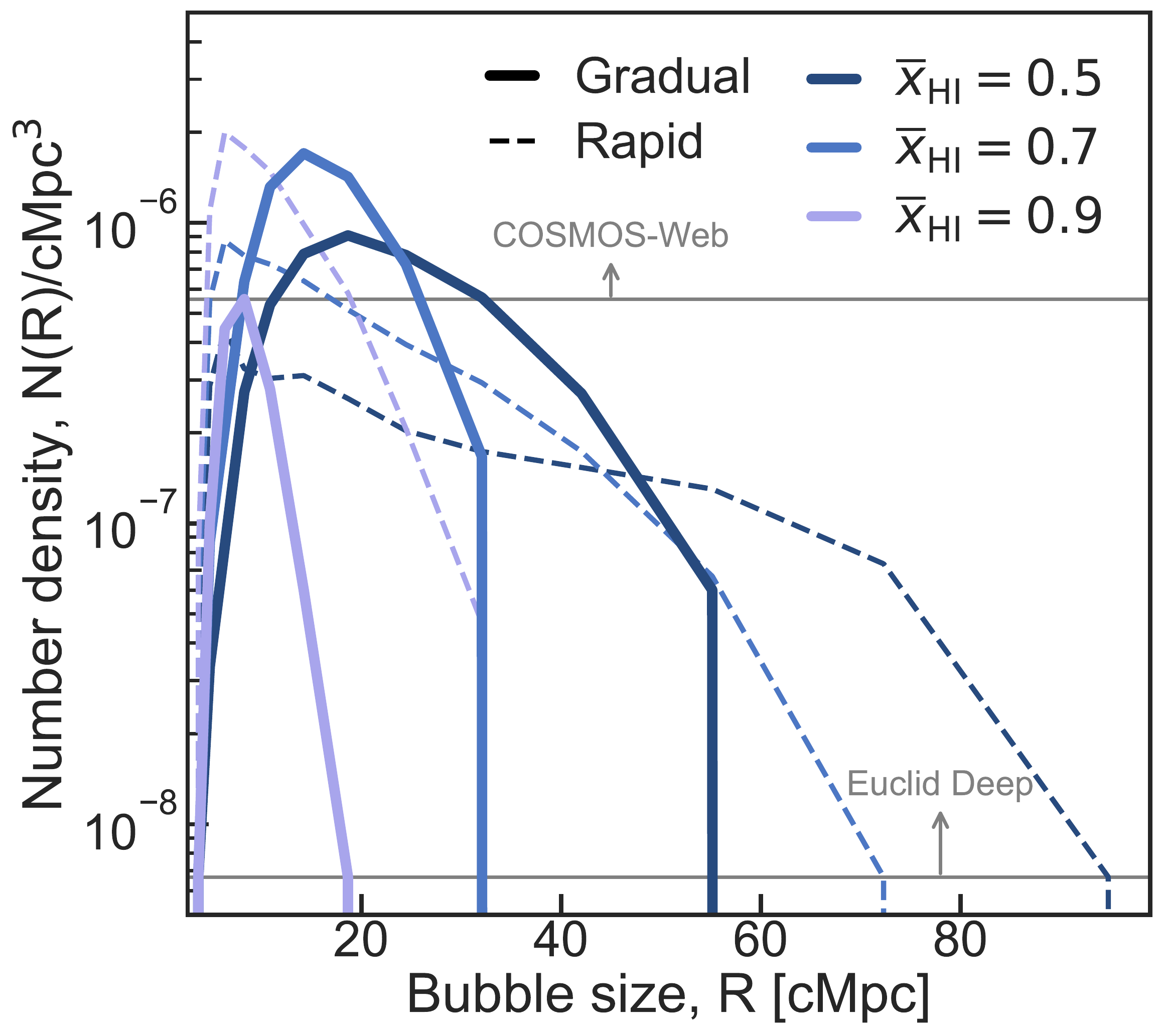}
    \caption{Number density of ionised bubbles for a range of \xHI and for our \FG\ (solid) and \BG\ models (dashed), calculated using the watershed algorithm. We show the inverse of the survey volume for COSMOS-Web (120 cMpc)$^3$ and Euclid Deep (530 cMpc)$^3$ as horizontal lines, marking the number density where one bubble is expected in that volume.}
    \label{fig:forecast_BSF_euc}
\end{figure}

In Figure~\ref{fig:forecast_BSF_euc} we plot our predicted `bubble size function' down to this resolution limit: the number density of ionised bubbles as a function of bubble size, for our \FG\ and \BG\ model at $\xHI=[0.5,0.7,0.9]$. The number of bubbles with $R\simgt10$\,cMpc can be considered a proxy for a cluster of \lya-emitting galaxies, as $\sim30-50\%$ of \lya flux should be transmitted through regions this large \citep{Mason2020}.
As the neutral fraction decreases, as above, we expect to find an increasing number of large ionised regions, and the number of small ionised regions decreases as bubbles overlap. Figure~\ref{fig:forecast_BSF_euc} again shows the clear difference in the predicted number and size of ionised regions for the different reionisation models, as discussed in Section~\ref{sec:results_BSD_model}.
We mark the survey volume of COSMOS-Web and Euclid Deep, (120cMpc)$^{3}$ and (530cMpc)$^{3}$ at $z=8$, respectively, as horizontal lines. The survey volume of the Roman High-Latitude survey (not shown) is (1816cMpc)$^{3}$.

We note that when $\xHI \simlt 0.7$ we expect a significant fraction of bubbles with $R\simgt50$\,cMpc (see Figures~\ref{fig:pR_Muv} and \ref{fig:PR_differentodxIH}). COSMOS-Web is thus unlikely to capture the full extent of large ionised bubbles. \citet{Kaur2020} demonstrated that a simulated volume of $>$(250 cMpc)$^3$ is required for convergence of the 21-cm power spectrum during reionisation, so it is likely that a similar volume must be observed to be able to robustly measure the bubble size distribution, thus we expect the Euclid Deep and Roman High Latitude surveys can robustly sample the full bubble size distribution. 

However, to detect $R>10$\,cMpc ionised bubbles, requires not only a large survey volume, but sufficient survey depth to detect the high redshift UV-bright galaxies which signpost large ionised regions. Only the Roman Space Telescope (RST) \citep{Akeson2019} is likely to be able to carry out bubble counting. \citet{Zackrisson2020} study the number of galaxies within a $\mathrm{V}_{\rm ion}=1000$ cMpc$^{3}$ bubble that can be detected with upcoming photometric surveys with instruments such as Euclid, JWST, and RST. They found that the Euclid Deep survey can barely detect one $\MUV\approx-21$ galaxy in that volume at $z>7$ given its detection limit, meaning that identifying large overdensities will be challenging. 
By contrast, a $\approx20$ deg$^{2}$ deep field observation by RST could detect $\sim10\,\MUV \simlt -18.5$ galaxies at $z=7-10$ in a $\mathrm{V}_{\rm ion}=1000$ cMpc$^{3}$ volume. The wide survey area ($\approx$ (400 cMpc)$^{3}$ at $z\sim8\pm0.2$) and survey depth of a RST deep field observation will allow us to identify the UV-bright galaxies which trace large ionised regions.
Deeper imaging or slitless spectroscopy around the UV-bright sources, for example with JWST to confirm overdensities, followed by \lya spectroscopy of these regions would enable estimates of the number density of ionised bubbles.

These results demonstrate that counting the number of overdensities of LAEs in a volume can be a useful estimate of the bubble size distribution, as they will probe ionised regions $\simgt1$\,pMpc, and thus \xHI (especially at $\xHI>0.5$).
For example, in our fiducial \FG\ model we expect no $R>10$\,cMpc bubbles in the COSMOS-Web volume when $\xHI=0.9$. This implies detection of clusters of LAEs in this volume at a given redshift would indicate $\xHI < 0.9$ (or a reionisation morphology similar to our \BG\ model). We expect tens of large bubbles in this volume when $\xHI < 0.7$. However, multiple sightline observations, for example, a counts-in-cells approach can be a more efficient tool to recover the distribution than a single area survey \citep[e.g.,][]{Mesinger2008a}.

\begin{figure}
    \centering
    \includegraphics[width=\columnwidth]{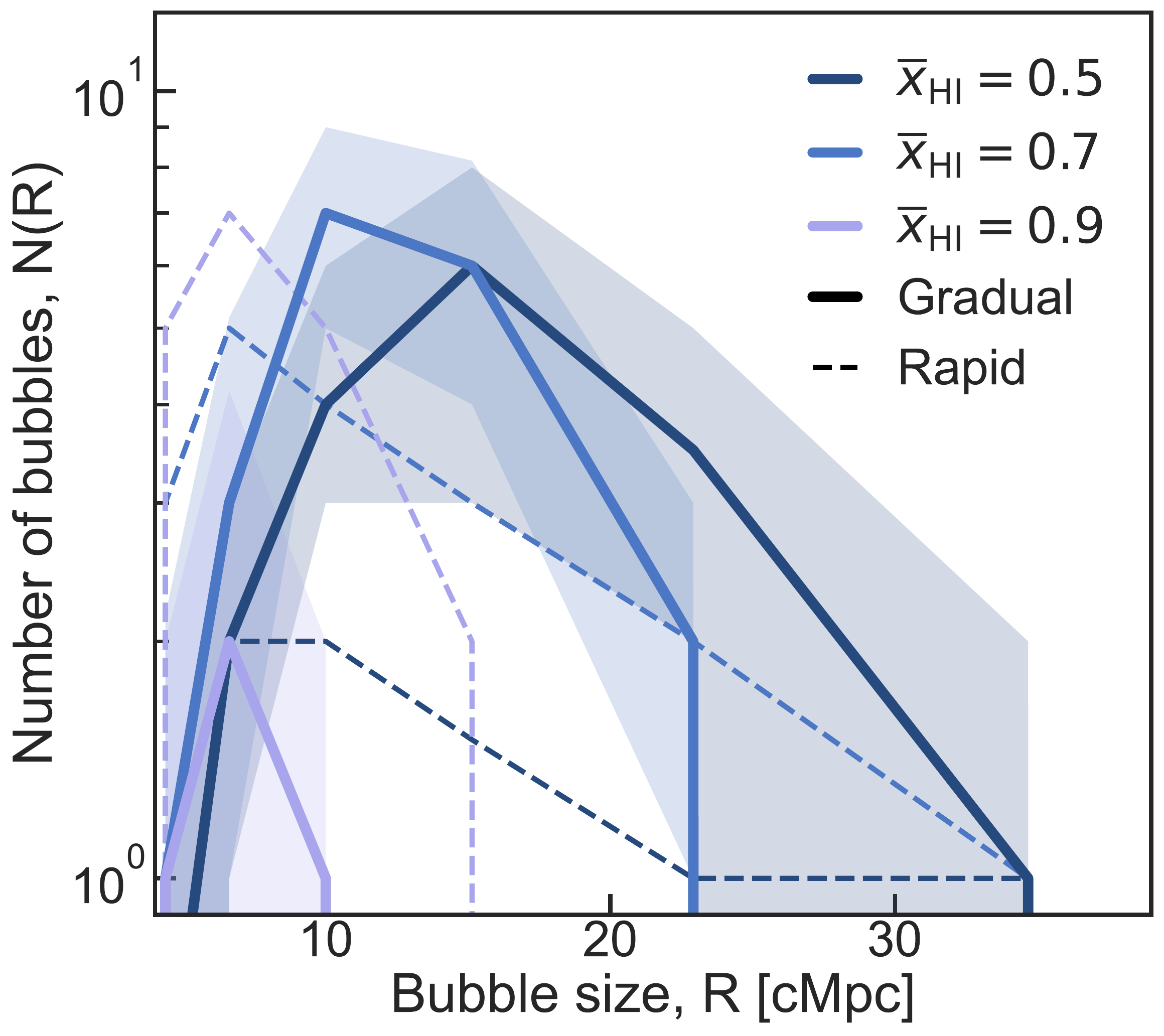}
    \caption{Number of ionised bubbles we predict from multiple realisations of a COSMOS-Web-like survey for our \FG\ and \BG\ models at a range of neutral fractions. The lines show the median number counts and the shaded regions are the 16-84 percentile of the number counts, demonstrating the large cosmic variance in this volume.}
    \label{fig:forecast_nbub_cos}
\end{figure}

However, bubble size functions in a COSMOS-Web-like survey volume have high cosmic variance, making it challenging to measure \xHI precisely. In Figure~\ref{fig:forecast_nbub_cos} we plot the median number of bubbles that can be observed by a COSMOS-Web-like survey using 50 realisations along with the 16-84 percentile number counts for our \FG\ model. The variance is large enough to make the bubble size functions at $\xHI=0.5-0.7$ indistinguishable. We do not plot the variance for the \BG\ model for clarity, but when taking that into account, we cannot discriminate between the bubble size functions of \FG\ and \BG\ with a COSMOS-Web-like survey. 

\section{Discussion}
\label{sec:disc}

In the following section we compare our results to those obtained from other simulations (Section~\ref{sec:disc_compare}) and discuss the implications of our results for the reionisation history and identifying the primary sources of reionisation (Section~\ref{sec:disc_reionisation}).

\subsection{Comparison to other simulations}
\label{sec:disc_compare}

\begin{figure}
    \centering
    \includegraphics[width=\columnwidth]{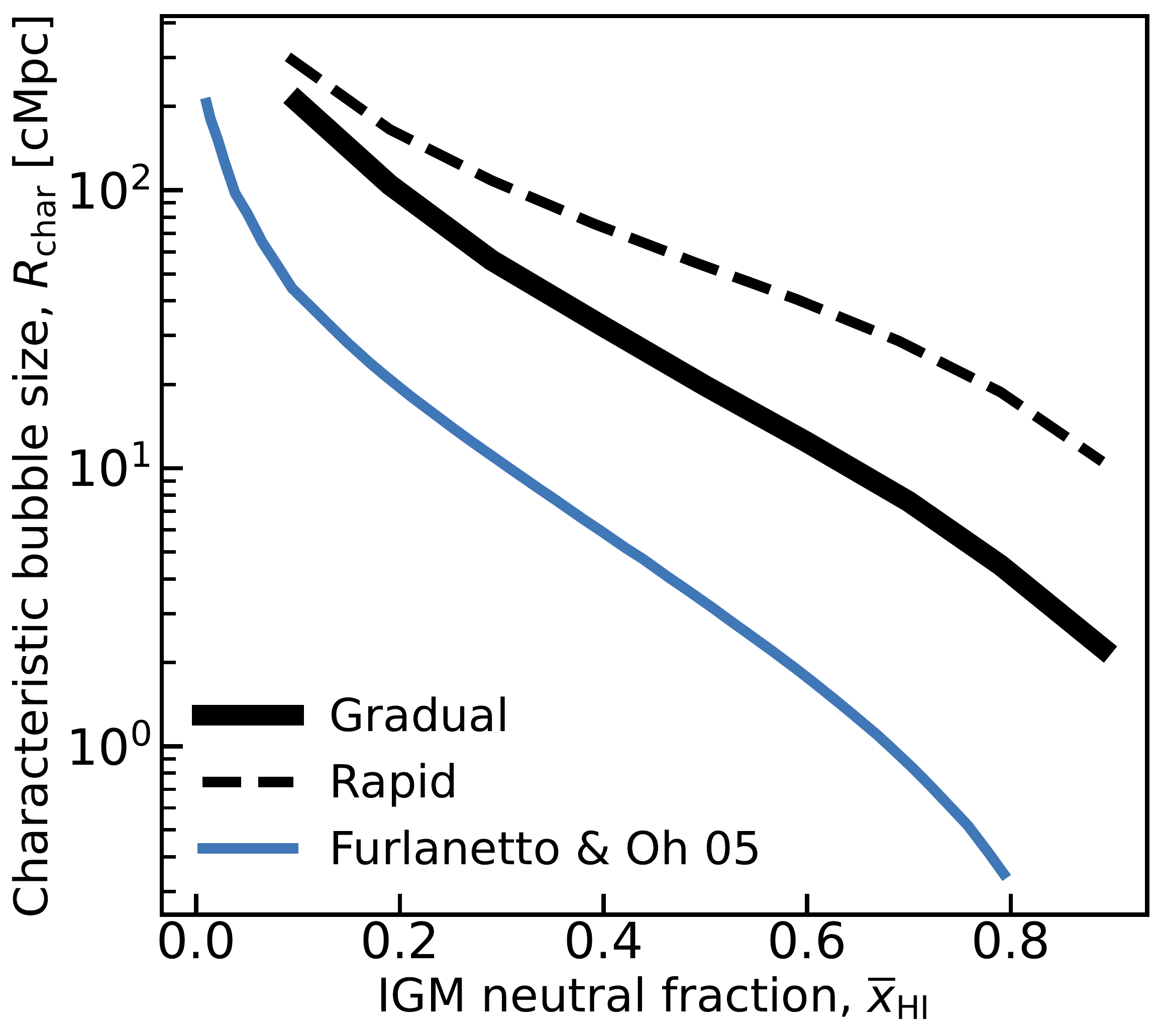}
    \caption{The evolution of `characteristic' bubble sizes as a function of \xHI for our simulations compared to previous work. We show the mean size of ionised regions in this work (black) for the \FG\ and \BG\ models (solid and dashed lines respectively), and the characteristic size of ionised region in \citet{Furlanetto2005} (blue). As discussed in Section~\ref{sec:results_BSD_model}, characteristic sizes of ionised regions in the \BG\ model are much larger than those in the \FG\ model at fixed \xHI. The excursion set formalism used by \citet{Furlanetto2005} can underestimate the sizes of ionised regions by over an order of magnitude as it does not account for overlapping regions.}
    \label{fig:cp_Rion_sim}
\end{figure}

In this work, we characterise the bubble size distributions around typically observed reionisation-era galaxies for the first time over the full timeline of reionisation. Previously, only the total bubble size distribution has been modelled as a function of the neutral fraction \citep[e.g.][]{Furlanetto2005,Mesinger2007,McQuinn2007b,Seiler2019a}. 

In principle, the full bubble size distribution measured in this work should agree with previous works of similar reionisation setups. 
However, as seen in Figure~\ref{fig:cp_Rion_sim} our mean bubble size is significantly larger than the characteristic bubble size modelled by \citet{Furlanetto2005}. Our bigger size comes from our use of the mean-free-path (MFP) method which is capable of taking into account the size of overlapped bubbles. The \cite{Furlanetto2005} model underestimates the typical bubble size because they calculate bubbles via the excursion set formalism: as found by \cite{Lin2016}, this method can underestimate bubble sizes by an order of magnitude. 
Our mean bubble size is comparable to those in works which use the mean free path approximation \citep[e.g.,][]{Mesinger2007,Seiler2019a}, modulo minor differences due to different assumptions for the ionising source population, as expected looking at the difference between our \FG\ and \BG\ models.

Other works have explored the correlation between  ionised bubble size and galaxy luminosity. For example,
\citet{Geil2017} and \citet{qin_dark-ages_2022} presented results from the DRAGONS simulation \citep{Poole2016}, finding more luminous galaxies are more likely to reside in large ionised bubbles, and that UV-faint galaxies have a large scatter in their host bubble size, consistent with our results in Section~\ref{sec:results_BSD_MUV}. However, these works only investigated a single redshift, IGM neutral fraction and reionising source model. Furthermore, the DRAGONS simulation is only (100 cMpc)$^3$, meaning that it does not contain large numbers of rare overdensities and UV-bright galaxies \citep[it contains only 2 galaxies as bright as GNz11,][]{Mutch2016a} and thus their predicted bubble sizes around UV-bright galaxies were subject to substantial Poisson noise. \citet{Yajima2018} presented a model for the sizes of ionised bubbles around galaxies by modelling cosmological Stromgren spheres around each galaxy \citep[e.g.,][]{Shapiro1987,Cen2000a}, finding more massive and highly star-forming galaxies (and therefore more luminous) lie in larger ionised bubbles than low mass galaxies. However, this model does not take into account the overlapping of ionised regions, which can happen very early during reionisation \citep[e.g.,][]{Lin2016} and thus their bubble sizes will be underestimated.

Our results in Section~\ref{sec:results_BSD_MUV} highlight the importance of considering the expected ionised bubble size as a function of \MUV calculated using the MFP method. Previous works which used the characteristic bubble size predicted by \citet{Furlanetto2005} will thus be underestimating the size of ionised bubbles around observed galaxies at fixed neutral fraction. \citet{jung_texas_2020} estimated the ionised bubble size required to explain the drop in \lya transmission in the GOODS-N field at $z\sim7.6$, and compared this bubble size to the \citet{Furlanetto2005} characteristic bubble size as a function of neutral fraction to estimate $\xHI \sim 0.49\pm0.19$. Our results in Figure~\ref{fig:cp_Rion_sim} imply that this approach will lead to an underestimate in \xHI. This likely explains the discrepancy between the neutral fraction estimated by \citet{jung_texas_2020} and that inferred by \citet{Bolan2022} ($\xHI = 0.83_{-0.11}^{+0.08}$) at a similar redshift, which was obtained by sampling sightlines in inhomogeneous IGM simulations.

\subsection{Implications for the reionisation history and identification of primary ionising sources}
\label{sec:disc_reionisation}

Our results demonstrate that the visibility of \lya\ emission at $z>8$ is unexpected given our consensus timeline for reionisation. The visibility of \lya\ therefore implies some missing aspect in our understanding of reionisation. 

As discussed in Section~\ref{sec:res_interp} there are three possibilities: (1) \xHI is lower than previously inferred; (2) reionisation is dominated by rarer sources providing larger, rarer bubbles; (3) these galaxies have high intrinsic \lya production \citep{Stark2017,Tang2023} and facilitated transmission in the IGM if the \lya flux is emitted redward of systemic \citep[e.g.,][]{Dijkstra2011,Mason2018b}. These scenarios should be testable with spectroscopic observations in the field of high redshift \lya-emitters. The most important first step is confirming if the large regions really are ionised. As the \lya damping wing due to nearby neutral gas strongly attenuates \lya close to systemic velocity, detecting \lya with high escape fraction (estimated from Balmer lines) and very low velocity offset would be a key test to infer if the sources lie in large ionised regions. The $z>8$ LAEs that have been detected so far have \lya velocity offset $>300$\,km s$^{-1}$ \citep{Tang2023,Bunker2023}, thus the large ionised regions cannot be confirmed, but spectroscopy of the fainter galaxies \citep[which are more likely to emit \lya closer to systemic velocity,][]{Prieto-Lyon2023b} in these overdensities could be used to confirm large bubbles. These observations are now possible with JWST/NIRSpec, which can also importantly spectroscopically confirm overdensities. Excitingly, recent observations have discovered strong \lya at low velocity offsets at $z>7$, implying large ionised regions \citep{Tang2023,Saxena2023}, and we will discuss quantitative constraints on the sizes of ionised regions in a future work.

We have also shown the bubble size distribution around observable galaxies depends on both the average IGM neutral fraction \xHI and the reionising source model. As the characteristic bubble size evolves strongly with \xHI (Figure~\ref{fig:cp_Rion_sim}), we may be able to constrain the reionisation history by simply counting overdensities of LAEs as a function of redshift.
\citet{trapp_lyman-alpha_2022} recently used observed overdensities of LAEs to place joint constraints on the IGM neutral fraction and underlying matter density of those regions. That work is complementary to our approach in that it demonstrates a strong link between the overdensity of a region and the expected size of the ionised region around an overdensity.

In Sections~\ref{sec:results_BSD_model} and ~\ref{sec:res_forecasts} we show that the reionising source models have a strong impact on the predicted number of 
galaxies in large ionised bubbles early in reionisation. Finding evidence for a high number density of large ionised regions ($\simgt10$\,cMpc) at high redshift would thus provide evidence for reionisation driven by rare, bright sources. However, it is clear from our work that characteristic bubble sizes in different reionisation models at different \xHI can be degenerate, so focusing purely on observing overdensities likely to reside ionised regions will not be able to break this degeneracy.

As seen clearly in Figure~\ref{fig:sim}, the \BG\ model is characterised by biased, isolated large bubbles, thus it is much more likely that galaxies outside of overdensities will still be in mostly neutral regions early in reionisation in this scenario (see Figure~\ref{fig:pR_od_BG_FG}).
Thus to measure \xHI and fully break the degeneracy between reionisation morphologies requires observing a range of environments over time during reionisation. For example, observing the \lya transmission from multiple sightlines to galaxies at different redshifts \citep[e.g.,][]{Mesinger2008a,mason_beacons_2018, Whitler2020, Bolan2022} and the 21-cm power spectrum as a function of redshift \citep[e.g.,][]{Furlanetto2004,Geil2016}.

\section{Conclusions} \label{sec:conc}

We have produced large-scale (1.6 Gpc)$^3$ simulations of the reionising IGM using \cmfast and explored the size distribution of ionised bubbles around observable galaxies. Our conclusions are as follows:
\begin{enumerate}
    \item Observable galaxies ($\MUV < -16$) and galaxy overdensities are much less likely to reside in neutral regions compared to regions at the mean density. This is because galaxies are the source of reionisation.
    \item The bubble size distribution around UV-bright ($\MUV < -20$) galaxies and strong galaxy overdensities is biased to larger characteristic sizes compared to those in the full volume.
    \item At all stages of reionisation we find a trend of increasing characteristic host bubble size and decreasing bubble size scatter with increasing UV luminosity and increasing overdensity.
    \item As shown by prior works, we find the bubble size distribution strongly depends on both the IGM neutral fraction and the reionising source model. The difference between these models is most apparent in the early stages of reionisation, $\xHI > 0.5$: if numerous faint galaxies drive reionisation, we expect a gradual reionisation with numerous small bubbles, whereas if bright galaxies drive reionisation we expect a more rapid process characterised by larger bubbles biased around only the most overdense regions, with sizes $>30$\,cMpc even in a 90\% neutral IGM.
    \item We use our simulations to interpret recent observations of galaxy overdensities detected with and without \lya emission at $z \gtrsim 7$. We find the probability of finding a large ionised region with $\Rbub>1$\,pMpc, capable of transmitting significant \lya flux, at $z\approx7-8$ is high ($\simgt40-93$\%) for large-scale galaxy overdensities, implying that \lya-emitting galaxies detected at these redshifts are very likely to be in large ionised regions.
    \item We find a very low probability of the $z\approx8.7$ association of \lya emitters in the EGS field and the $z=10.6$ galaxy GNz11, also detected with \lya emission, to be in a large ionised bubble ($\sim11\%$ and $\sim7\%$, respectively). The \lya detections at such a high redshift could be explained by either: a lower neutral fraction ($\xHI\lesssim0.8$) than previously inferred; or if UV bright galaxies drive reionisation, which would produce larger bubbles; or if the intrinsic \lya production in these galaxies is unusually high.
    \item We make forecasts for the number density of ionised bubbles as a function of bubble size expected in the JWST COSMOS-Web survey and the Euclid Deep survey. Our fiducial model predicts no ionised regions $>10$\,cMpc in the COSMOS-Web volume unless $\xHI < 0.9$, with tens of large bubbles expected by $\xHI < 0.7$, though with large cosmic variance. We find Euclid and Roman wide-area surveys will have sufficient volume to cover the size distribution of ionised regions with minimal cosmic variance and should be able to detect the UV-bright galaxies which signpost overdensities. Deeper photometric and spectroscopic follow-up around UV-bright galaxies in these surveys to confirm overdensities and \lya emission could be used to infer \xHI and discriminate between reionisation models.
\end{enumerate}

Our simulations show that in interpreting observations of $z>6$ galaxies it is important to consider the galaxy environment. We showed the bubble size distribution around observable galaxies and galaxy overdensities can be significantly shifted from the bubble size distribution over the whole cosmic volume. This motivates using realistic inhomogeneous reionisation simulations, or at least tailored bubble size distributions to interpret observations.

Our results imply that the early stages of reionisation are still very uncertain. Identifying and confirming large ionised regions at very high redshift is a first step to understanding these early stages, and thus the onset of star formation. This is now possible with deep JWST/NIRSpec observations which could map the regions around $z>8$ \lya emitters. The detection of \lya with high escape fraction and low velocity offset from other galaxies in the observed $z>8$ overdensities could confirm whether the \lya\ emitters at $z>8$ are tracing unexpectedly large ionised regions \citep[e.g.,][]{Tang2023,Saxena2023}.

\section*{Acknowledgments}

TYL, CAM and AH acknowledge support by the VILLUM FONDEN under grant 37459. The Cosmic Dawn Center (DAWN) is funded by the Danish National Research Foundation under grant DNRF140. This work has been performed using the Danish National Life Science Supercomputing Center, Computerome. Part of this research was supported by the Australian Research Council Centre of Excellence for All Sky Astrophysics in 3 Dimensions (ASTRO 3D), through project \#CE170100013.

\section*{Data Availability}
Tables of bubble sizes as a function of \xHI, \MUV, and galaxy overdensity (using the same volume as in Section~\ref{sec:results_BSD_overdensity}) are publicly available here: \url{https://github.com/ting-yi-lu/bubble_size_overdensities_paper}. \\
Bubble size distributions around other overdensities can be distributed upon reasonable request to the authors.



\bibliographystyle{mnras}
\bibliography{library} 



\appendix

\section{Model UV luminosity function} \label{app:LF}

\begin{figure}
    \centering
    \includegraphics[width=\columnwidth]{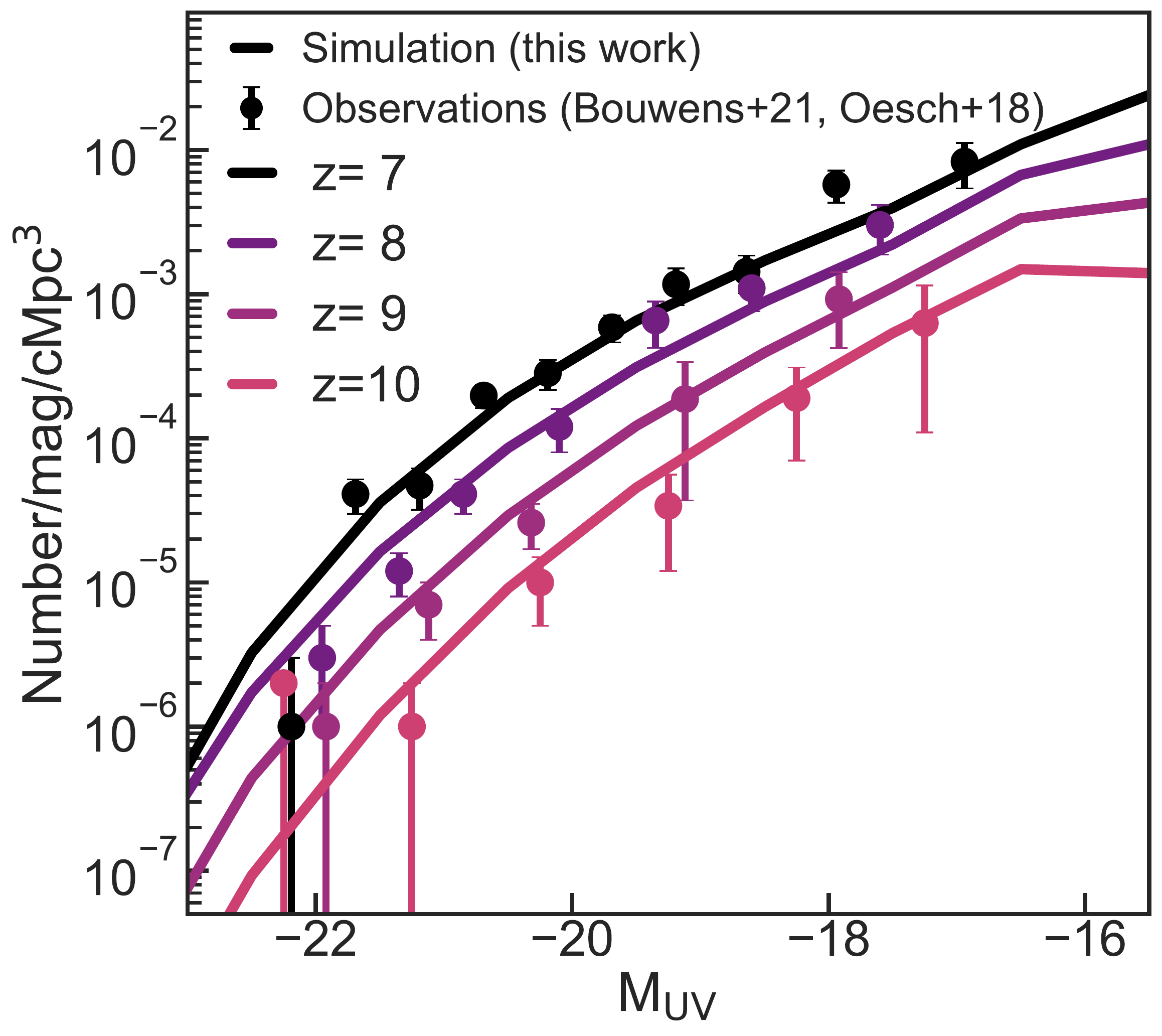}
    \caption{UV luminosity functions from our model at $z\sim7,8,9$ (solid lines) in comparison to HST measurements of the UV LF by \citet{Bouwens2021} ($z=7-9$) and \citet{Oesch2018a} ($z=10$).}
    \label{fig:LF}
\end{figure}

In Figure~\ref{fig:LF} we demonstrate that our model for assigning UV magnitudes to simulated halos (Section~\ref{sec:methods_galaxy}) reproduces observed UV luminosity functions over $z\sim7-10$ as required for our study. We note the apparent turnover at $\MUV \simgt -16$ is not physical, but arises due to enforcing a halo mass cut-off at $M_{\rm halo}=5\times 10^{9} M_{\odot}$ in our catalogue due to memory restrictions.

\section{Bubble size distributions at different redshifts} \label{app:z}

\begin{figure}
    \centering
    \includegraphics[width=\columnwidth]{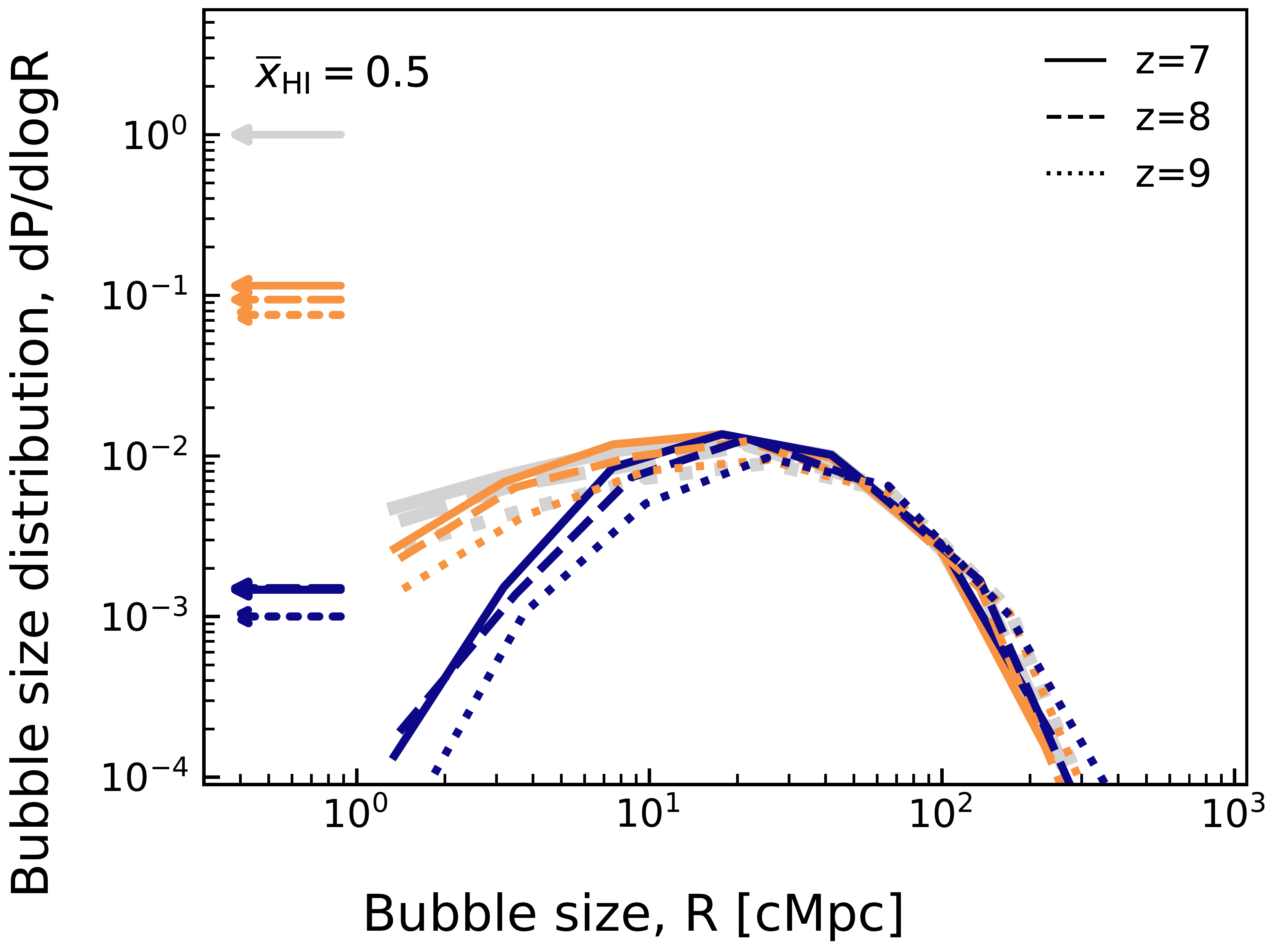}
    \caption{Bubble size distributions, \dpdlogR, as a function of redshift for $z=7$ (solid lines) and $z=8$ (dashed lines) and $z=9$ (dotted lines) for $\MUV=-16, -22$ at $\xHI=0.5$. We also show the total bubble size distribution as a thick grey line in each redshift. We see a minimal difference at different redshifts, but higher redshifts show slighter higher bubble sizes as we discuss in Appendix~\ref{app:z}.}
    \label{fig:pR_Muv_z7_z8}
\end{figure}

Here we compare the bubble size distributions at $z=7-9$. We show the bubble size distributions at $\xHI=0.5$ for each redshift in Figure~\ref{fig:pR_Muv_z7_z8}.  We see negligible differences as a function of redshift, but that bubble sizes are slightly larger at fixed neutral fraction at higher redshift. This is because we use a fixed halo mass cut-off to calculate the ionising emissivity (as described in Section~\ref{sec:methods_sims}), and at higher redshifts the same mass halo will be more biased, resulting in rarer, larger bubbles at fixed \xHI as in our \BG\ model. However, the difference between the bias of halos of fixed mass and different redshifts is much lower than the difference between the bias due to our two mass thresholds for the \FG\ and \BG\ model, so this redshift effect is minimal.

\section{Bubble size distribution models for observed overdensities} 
\label{app:obs}

\begin{figure*}
    \centering
    \includegraphics[width=0.8\textwidth]{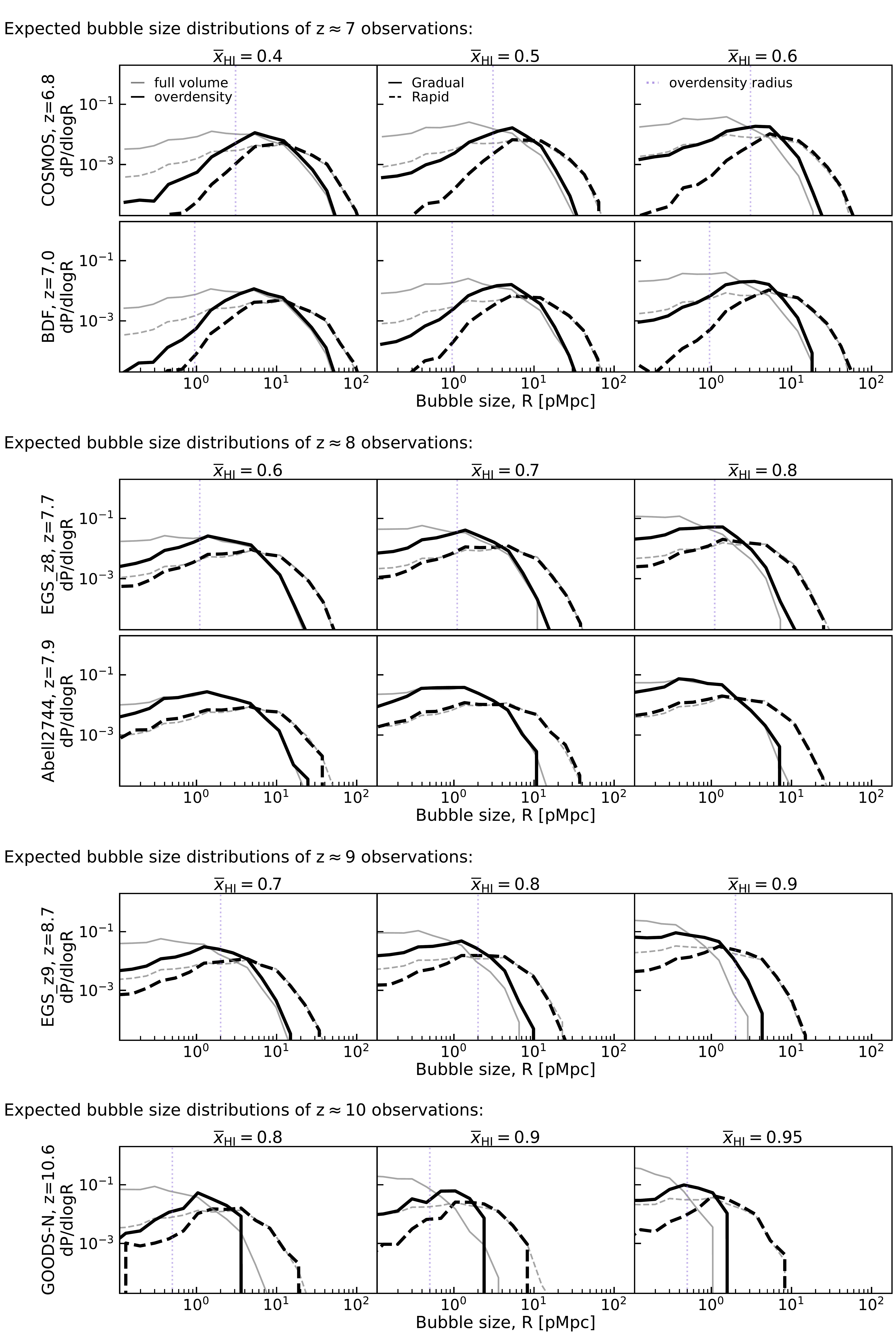}
    \caption{Bubble size distributions for the same overdensity observation setups as the COSMOS (top panel), BDF (second panel), EGS\_z8 (third panel), Abell2744 (fourth panel), EGS\_z9 (fifth panel), and GOODS-N (bottom panel), at the IGM neutral fractions expected at these redshifts \citep{Mason2019b}, from the \FG\ (solid) and \BG\ (dashed) models. The bubble size estimated by previous works \citep{Endsley2022, Castellano2016b, leonova_prevalence_2022, Morishita2022b,Jung2022, Tacchella2023} are marked with purple vertical lines. A summary of our simulation setup is given in Table~\ref{tab:bubbles}.}
    \label{fig:BSD_obs_all}
\end{figure*}

In Figure~\ref{fig:BSD_obs_all} we show the bubble size distribution for the observed overdensities described in Section~\ref{sec:res_interp}. In all plots we show the bubble size distribution in the full volume in the neutral fraction range expected given current constraints on reionisation \citep{Mason2019b}, and the bubble size distribution in regions as overdense as those observed.

\bsp	
\label{lastpage}
\end{document}

%% file: macros.tex
\newcommand{\vdag}{(v)^\dagger}
\newcommand{\myemail}{skywalker@galaxy.far.far.away}

\newcommand{\simgt}{\,\rlap{\lower 3.5 pt \hbox{$\mathchar \sim$}} \raise 1pt \hbox {$>$}\,}
\newcommand{\simlt}{\,\rlap{\lower 3.5 pt \hbox{$\mathchar \sim$}} \raise 1pt \hbox {$<$}\,}
\newcommand{\dd}{\mathrm{d}}

\newcommand{\BE}{\begin{equation}}
\newcommand{\EE}{\end{equation}}
\newcommand{\BEA}{\begin{eqnarray}}
\newcommand{\EEA}{\end{eqnarray}}

\newcommand{\Ob}{\Omega_\textrm{b}}
\newcommand{\Om}{\Omega_\textrm{m}}
\newcommand{\OL}{\Omega_\Lambda}
\newcommand{\rhoc}{\rho_\textrm{c}}

\newcommand{\CHII}{C_\textrm{\ion{H}{2}}}
\newcommand{\DV}{\ifmmode{\Delta v}\else $\Delta v$\xspace\fi}

\newcommand{\Tigm}{{\mathcal{T}_\textsc{igm}}}

\newcommand{\HI}{\ifmmode{\textsc{hi}}\else H\textsc{i}\fi\xspace}
\newcommand{\HII}{\ifmmode{\textsc{hii}}\else H\textsc{ii}\fi\xspace}
\newcommand{\OII}{[O\textsc{ii}]}
\newcommand{\OIII}{O\textsc{iii}}
\newcommand{\CIV}{C\textsc{iv}}
\newcommand{\HeII}{He\textsc{ii}}
\newcommand{\Ha}{\ifmmode{\mathrm{H}\alpha}\else H$\alpha$\fi\xspace}
\newcommand{\OiiiHb}{\ifmmode{\mathrm{[OIII]+H}\beta}\else [OIII]+H$\beta$\fi\xspace}

\newcommand{\Msun}{\ifmmode{M_\odot}\else $M_\odot$\xspace\fi}
\newcommand{\MUV}{\ifmmode{M_\textsc{uv}}\else $M_\textsc{uv}$\xspace\fi}
\newcommand{\fesc}{\ifmmode{f_\mathrm{esc}}\else $f_\mathrm{esc}$\xspace\fi}
\newcommand{\lya}{\ifmmode{\mathrm{Ly}\alpha}\else Ly$\alpha$\xspace\fi}
\newcommand{\vrot}{v_\textrm{rot}}

\newcommand{\nh}[1][]{\ifmmode{\overline{n}_\textsc{h}^{#1}}\else $\overline{n}_\textsc{h}$\xspace\fi}

\newcommand{\Wobs}{\ifmmode{W_\textrm{obs}}\else $W_\textrm{obs}$\xspace\fi}

\newcommand{\xHI}{\ifmmode{\overline{x}_\HI}\else $\overline{x}_\HI$\xspace\fi}
\newcommand{\trec}{\ifmmode{t_\textrm{rec}}\else $t_\textrm{rec}$\xspace\fi}
\newcommand{\clump}[1][]{\ifmmode{C_\HII^{#1}}\else $C_\HII$\xspace\fi}
\newcommand{\xiion}{\ifmmode{\xi_\mathrm{ion}}\else $\xi_\mathrm{ion}$\xspace\fi}

\newcommand{\Nion}{\ifmmode{\dot{N}_{\mathrm{ion}}}\else $\dot{N}_\mathrm{ion}$\xspace\fi}
\newcommand{\Rion}[1][]{\ifmmode{R_\mathrm{ion}^{#1}} \else $R_\mathrm{ion}$\xspace\fi}

\newcommand{\Rchar}{\ifmmode{R_{\mathrm{char}}}\else $R_\mathrm{char}$\xspace\fi}
\newcommand{\Rbub}{\ifmmode{R_{\mathrm{ion}}}\else $R_\mathrm{ion}$\xspace\fi}
\newcommand{\Rexp}{\ifmmode{\langle R \rangle}\else $\langle R \rangle$\xspace\fi}
\newcommand{\dpdlogR}{\ifmmode{dp/d\log_{10}R}\else $dp/d\log_{10}R$\xspace\fi}
\newcommand{\od}{\ifmmode{N/\langle N \rangle}\else $N/\langle N \rangle$\xspace\fi}

\newcommand{\fdens}{\,erg s$^{-1}$ cm$^{-2}$\xspace}
\newcommand{\kms}{\,\ifmmode{\mathrm{km}\,\mathrm{s}^{-1}}\else km\,s${}^{-1}$\fi\xspace}
\newcommand{\cm}{\,\ifmmode{\mathrm{cm}}\else cm\fi\xspace}

\newcommand{\HST}{\textit{HST}}
\newcommand{\JWST}{\textit{JWST}}
\newcommand{\WFIRST}{\textit{WFIRST}}

\newcommand{\cmfast}{\texttt{21cmfast}\xspace}
\newcommand{\BG}{\textit{Rapid}}
\newcommand{\FG}{\textit{Gradual}}
\newcommand{\Rres}{1\,cMpc}

\newcommand{\NB}[1]{\textbf{\color{red} #1}}
\newcommand{\tnm}[1]{$^\textrm{#1}$}